\documentclass[12pt]{article}

\usepackage[totalwidth=470truept,totalheight=636truept]{geometry}
\usepackage{latexsym,graphicx,amsfonts,amssymb,amsmath,color, cancel}
\usepackage[hypertexnames=false,hidelinks]{hyperref}

\def\theequation{\arabic{section}.\arabic{equation}}

\renewcommand{\theequation}{\thesection.\arabic{equation}}
\global\arraycolsep=1truept
\renewcommand{\theequation}{\arabic{section}.\arabic{equation}}

\begin{document}

\null

\vskip1truecm

\begin{center}
{\huge \textbf{Renormalization-Group Techniques}}

\vskip.8truecm

{\huge \textbf{for Single-Field Inflation}}

\vskip.8truecm

{\huge \textbf{in Primordial Cosmology}}

\vskip.8truecm

{\huge \textbf{and Quantum Gravity}}

\vskip1truecm

\textsl{Damiano Anselmi}$^{1,2,3,a}$\textsl{, Filippo Fruzza}$^{4,b}$\textsl{%
\ and Marco Piva}$^{3,c}$

\vskip .1truecm

$^{1}${\scriptsize \textit{Dipartimento di Fisica \textquotedblleft Enrico
Fermi\textquotedblright , Universit\`{a} di Pisa, Largo B. Pontecorvo 3,
56127 Pisa, Italy}}

$^{2}${\scriptsize \textit{INFN, Sezione di Pisa, Largo B. Pontecorvo 3,
56127 Pisa, Italy}}

$^{3}${\scriptsize \textit{National Institute of Chemical Physics and
Biophysics, R\"{a}vala 10, Tallinn 10143, Estonia}}

$^{4}${\scriptsize \textit{Via Alfieri 10, San Giuliano Terme, 56017, Pisa,
Italy,}}

$^{a}$ {\footnotesize damiano.anselmi@unipi.it, } $^{b}$ {\footnotesize %
filippo.fruzza@live.it, } $^{c}$ {\footnotesize marco.piva@kbfi.ee }

\vskip1truecm

\textbf{Abstract}
\end{center}

We study inflation as a \textquotedblleft cosmic\textquotedblright\
renormalization-group flow. The flow, which encodes the dependence on the
background metric, is described by a running coupling $\alpha $, which
parametrizes the slow roll, a de Sitter free, analytic beta function and
perturbation spectra that are RG invariant in the superhorizon limit. Using
RG invariance as a guiding principle, we classify the main types of flows
according to the properties of their spectra, without referring to their
origins from specific actions or models. Novel features include spectra with
essential singularities in $\alpha $ and violations of the relation $r+8n_{%
\text{t}}=0$ to the leading order. Various classes of potentials studied in
the literature can be described by means of the RG approach, even when the
action includes a Weyl-squared term, while others are left out. In known
cases, the classification helps identify the models that are ruled out by
data. The RG approach is also able to generate spectra that cannot be
derived from standard Lagrangian formulations.

\vfill\eject

\section{Introduction}

\label{intro}\setcounter{equation}{0}

Inflation explains the approximate isotropy and homogeneity of the cosmic
background radiation by means of a primordial accelerated expansion \cite%
{englert,starobinsky,kazanas,sato,guth,linde,steinhardt,linde2}. The quantum
fluctuations are sources of the perturbations that originate the present
large-scale structure of the universe \cite%
{mukh,mukh2,hawk,guth2,staro2,bardeen,mukh3}.

The expansion of the universe can be driven by scalar fields rolling down a
potential \cite{reviews}. Various scenarios of this type lead to a scalar
perturbation spectrum that is compatible with observations \cite%
{encicl,Planck18}. In a \textquotedblleft geometric\textquotedblright\
approach, instead, inflation is driven by gravity itself, as in the
Starobinsky $R+R^{2}$ model \cite{starobinsky} and the $f(R)$ theories \cite%
{defelice,otherfR}. A third approach is to study inflation as a
\textquotedblleft cosmic\textquotedblright\ renormalization-group (RG) flow 
\cite{CMBrunning,FakeOnScalar}, which is generated by the dependence on the
background metric. The spectra of the cosmological perturbations satisfy
equations of the Callan-Symanzik type in the superhorizon limit.

The RG flow of quantum field theory and the one of inflationary cosmology
have different origins, but many common features. The former is due to
ultraviolet divergences (in flat space), the latter is due to the nontrivial
dependence on the background FLRW metric. The sliding scale $\mu $ is mapped
onto the conformal time $\tau $, while the roles of the couplings are played
by the slow roll parameters. The correlation functions are the perturbation
spectra and the Callan-Symanzik equation becomes the conservation on
superhorizon scales. Asymptotic freedom becomes the de Sitter limit, etc.

It is convenient to summarize the present status of the correspondence in
the following table, which also provides a useful vocabulary:%
\begin{equation*}
\begin{tabular}{rcl}
Quantum field theory &  & Inflationary cosmology \\ 
RG\ flow & \multicolumn{1}{l}{$\leftrightarrow $} & slow roll \\ 
couplings $\alpha $, $\lambda $ ... & \multicolumn{1}{l}{$\leftrightarrow $}
& slow-roll parameters $\epsilon $, $\delta $ ... \\ 
beta functions & \multicolumn{1}{l}{$\leftrightarrow $} & equations of $a(t)$%
, $H(t)$ ... \\ 
sliding scale $\mu $ & \multicolumn{1}{l}{$\leftrightarrow $} & conformal
time $\tau $ (or $\eta =-k\tau $) \\ 
correlation functions & \multicolumn{1}{l}{$\leftrightarrow $} & 
perturbation spectra \\ 
Callan-Symanzik equation & \multicolumn{1}{l}{$\leftrightarrow $} & RG
equation at superhorizon scales \\ 
RG\ invariance & \multicolumn{1}{l}{$\leftrightarrow $} & conservation on
superhorizon scales \\ 
asymptotic freedom & \multicolumn{1}{l}{$\leftrightarrow $} & de Sitter
limit in the infinite past \\ 
subtraction scheme & \multicolumn{1}{l}{$\leftrightarrow $} & Einstein
frame, Jordan frame, etc. \\ 
dimensional transmutation & \multicolumn{1}{l}{$\rightarrow $} & $\tau $
drops out from the spectra, \textquotedblleft replaced\textquotedblright\ by 
$k$ \\ 
running coupling & \multicolumn{1}{l}{$\rightarrow $} & ok \\ 
resummation of leading logs & \multicolumn{1}{l}{$\rightarrow $} & ok \\ 
?? & \multicolumn{1}{l}{$\leftarrow $} & potential $V(\phi )$ \\ 
anomalous dimensions & \multicolumn{1}{l}{$\rightarrow $} & 0%
\end{tabular}%
\end{equation*}%
Note that we have no analogue of the anomalous dimensions, as far as we know
now. Switching from the Einstein frame to the Jordan frame can be seen as a
scheme change in the QFT language.

The correspondence is useful to enhance the calculations of the spectra to
higher orders with little effort, by means of RG techniques imported from
quantum field theory, like the use of running couplings and the resummation
of leading logs. Although we cannot test those corrections experimentally in
a foreseeable future, the gain in our undestanding is important. However,
several aspects of the correspondence are awaiting to be clarified. How deep
can we push the correspondence between the two types of flows? Can we
describe every potential $V(\phi )$ using the language of quantum field
theory? Are there RG flows that cannot be described by means of potentials?
Can we include $f(R)$ theories in the RG approach to inflationary cosmology?
In this paper we address some of these issues.

Let us anticipate some answers we find. First of all, we learn that we
cannot describe every potential $V(\phi )$ in the RG language. Actually,
certain classes of potentials fit better than others. Second, most $f(R)$
theories are still unreachable from the RG point of view. Third, the RG
setting provides a path to a more \textquotedblleft
axiomatic\textquotedblright\ approach, solely based on the properties of
perturbation spectra, derived from a Mukhanov-Sasaki action rather than a
modified gravity Lagrangian. An axiomatic approach may allow us to explore
scenarions following from alternative approaches to quantum gravity (like
loop quantum gravity or string theory). Among the other things, it allows us
to evade the hypotheses that imply the consistency relation $r+8n_{\text{t}%
}\simeq 0$.

In this paper we focus on single-field inflation. For a treatment of these
issues in double-field inflation, see ref. \cite{double}.

In the axiomatic spirit just mentioned, we demand that

$a$) the beta function is perturbative in the coupling $\alpha $, starts
quadratically in $\alpha $ and is de Sitter free in the infinite past (i.e., 
$\alpha =0$ gives the de Sitter background for $t\rightarrow -\infty $);

$b$) the power spectra are perturbative, apart from overall factors, and RG
invariant in the superhorizon limit.

Requirement $a$) implies that the beta function must behave like the one of
an asymptotically free quantum field theory, such as QCD, in the
perturbative region. The overall factors mentioned in point $b$) can be
negative or fractional powers of $\alpha $, or even essential singularities.
Negative powers are already familiar from the scalar spectra.

Defining the coupling $\alpha $ as $\sqrt{4\pi G/3}\dot{\phi}/H$, where $%
\phi $ is the inflaton and $H$ is the Hubble parameter, we can describe a
large number of potentials studied in the literature by means of RG flows.
However, many others are left out. To cover larger classes of models, it
might be necessary to satisfy requirements $a$) and $b$) with a different
definition of coupling, related to $\alpha $ in a non perturbative way. This
possibility is not explored here. On the other hand, the approach based on
the cosmic RG flow allows us to study cases that cannot be described by
means of more common approaches.

The strategy is to start from a general Mukhanov-Sasaki action%
\begin{equation}
S_{\text{MS}}=\frac{1}{2}\int \mathrm{d}\eta \left[ w^{\prime 2}-\left(
1+\Delta h\right) w^{2}+\frac{2+\sigma }{\eta ^{2}}w^{2}\right] ,  \label{sm}
\end{equation}%
for some perturbation $w(\eta )$, where $\eta $ is a rescaled conformal
time. Here $\Delta h$ and $\sigma $ are analytic functions of $\alpha $ that
vanish for $\alpha =0$. We refer to $\Delta h$ \ as a \textquotedblleft mass
renormalization\textquotedblright , which affects how we impose the
Bunch-Davies vacuum condition. We do not need to be specific about the
theory or model that gives (\ref{sm}). We can build the spectra associated
with (\ref{sm}) by solely requiring that they be RG invariant in the
superhorizon limit. This property is sufficient to determine them up to an $%
\alpha $-independent normalization constant, which can be fixed in the de
Sitter limit $\alpha =0$.

In several cases, the spectra exhibit essential singularities in $\alpha $.
Moreover, they often violate the \textquotedblleft consistency
condition\textquotedblright\ $r+8n_{\text{t}}=0$ \cite{peters}, which is
known to hold to the leading order in single-field slow-roll models,
independently of the scalar potential $V(\phi )$. There is no contraddiction
with the literature, because our \textquotedblleft
axiomatic\textquotedblright\ approach allows us to evade the assumptions
that imply the consistency condition just mentioned. This suggests that the
cosmic RG approach is essentially different from the other approaches,
although it intersects them in a number of cases.

The Starobinsky $R+R^{2}$ model works well phenomenologically. However, once
we add $R^{2}$ it is worth to consider the inclusion of the square $C_{\mu
\nu \rho \sigma }C^{\mu \nu \rho \sigma }$ $\equiv $ $C^{2}$ of the Weyl
tensor $C_{\mu \nu \rho \sigma }$ as well, because $C^{2}$ and $R^{2}$ have
the same dimensions in units of mass. So doing, we obtain the Lagrangian $%
R+R^{2}+C^{2}$, which is renormalizable (once the cosmological constant is
switched on), but leads to a theory plagued by ghosts, if quantized by means
of the standard methods \cite{stelle}. The problem of ghosts can be overcome
by formulating a physically different theory \cite{LWgrav,UVQG,Absograv} in
terms of purely virtual particles, or fakeons \cite{fakeons}. The properties
of the new theory can be appreciated particularly well in primordial
cosmology \cite{ABP}, where, due to a bound relating the coefficients of $%
C^{2}$ and $R^{2}$, which must be satisfied to make the formulation in terms
of fake particles consistent, the physics changes even on superhorizon
scales \cite{ABP}. The main outcome is a stringent prediction for the
tensor-to-scalar ratio $r$, not available in other formulations. The models
with $C^{2}$ provide an interesting arena for theoretical investigations,
since they are the only ones known to date which lead to a nontrivial mass
renormalization $\Delta h$.

We stress that in order to be able to talk about renormalization-group flow,
it is not enough to identify a flow that is governed by an autonomous first
order differential equation%
\begin{equation}
\frac{\mathrm{d}x}{\mathrm{d}t}=f(x).  \label{auto}
\end{equation}%
It is not even sufficient to show that there exists an equilibrium point $%
x_{0}$ ($f(x_{0})=0$) that is Lyapunov stable (i.e., such that the solutions
that start close enough to $x_{0}$ remain close enough forever), or one that
is asymptotically stable (i.e., such that the solutions that start close
enough to $x_{0}$ remain close enough to $x_{0}$ and converge to $x_{0}$).
As realized in ref.s \cite{CMBrunning,FakeOnScalar}, the key ingredient is
having correlation functions (the perturbation spectra, in our case) that
satisfy equations of the Callan-Symanzik type, $f(x)$ being the beta
function for the coupling $x$.

Note that the RG techniques studied here and in ref.s \cite%
{CMBrunning,FakeOnScalar} are essentially different from the
\textquotedblleft beta function formalism\textquotedblright\ studied in \cite%
{pieroni}. The autonomous differential equation associated with the first
class of models was discussed in \cite{kiritsis} in the context of
holography, but not related to the RG properties of the correlation
functions and spectra. Among earlier studies on the running behaviors of the
spectral indices we mention those of refs. \cite{run1}. Calculations of
subleading corrections to the perturbation spectra (in models without
fakeons and without $C^{2}$) are done in refs. \cite{run2}.

The paper is organized as follows. In section \ref{betaf} we recall the main
properties of the cosmic RG flow and show how it originates from specific
actions. In section \ref{flowtopot} we classify the flows by ignoring their
origins from specific actions or models. We also relate them to classes of
known and less known potentials, when possible. In section \ref{spectralRG}
we build the perturbation spectra in this general setting, by means of RG
invariance. In section \ref{specI} we calculate the spectra for the classes
of flows identified in section \ref{flowtopot}, in the absence of a mass
renormalization $\Delta h$. In section \ref{acttospec} we apply the results
to the RG flows associated with standard actions. In section \ref{pottoflow}
we investigate the reversed approach, which means search for the flow
associated with a given potential. We also list the cases that we cannot
treat at the moment. In section \ref{massren} we extend the results to the
flows with a nontrivial $\Delta h$ and the presence of purely virtual
particles. Section \ref{conclusions} contains the conclusions and appendix %
\ref{formulas} collects reference formulas and higher-order corrections.

\section{Inflationary beta function and cosmic RG\ flow}

\label{betaf}\setcounter{equation}{0}

In this section we recall the formulation of inflation as an RG\ flow. We
call it \textquotedblleft cosmic\textquotedblright\ RG\ flow to emphasize
that it is originated by the dependence on the background metric and not by
the radiative corrections.

The starting point is to consider actions such as%
\begin{equation}
S=-\frac{1}{16\pi G}\int \mathrm{d}^{4}x\sqrt{-g}R+\frac{1}{2}\int \mathrm{d}%
^{4}x\sqrt{-g}\left( D_{\mu }\phi D^{\mu }\phi -2V(\phi )\right) ,
\label{action}
\end{equation}%
or%
\begin{equation}
S=-\frac{1}{16\pi G}\int \mathrm{d}^{4}x\sqrt{-g}\left( R+\frac{1}{2m_{\chi
}^{2}}C_{\mu \nu \rho \sigma }C^{\mu \nu \rho \sigma }\right) +\frac{1}{2}%
\int \mathrm{d}^{4}x\sqrt{-g}\left( D_{\mu }\phi D^{\mu }\phi -2V(\phi
)\right) ,  \label{sqgeq}
\end{equation}%
where $V(\phi )$ is an arbitrary potential. For convenience, the
cosmological term is switched off. We do not add a term proportional to $%
R^{2}$ to (\ref{sqgeq}), because it would lead to models of double-scalar
inflation, which are beyond the scope of this paper. Yet, the theory $%
R+R^{2}+C^{2}$ is equivalent to a particular case of (\ref{sqgeq}) when $%
V(\phi )$ is the Starobinsky potential.

The FLRW metric%
\begin{equation}
g_{\mu \nu }=\text{diag}(1,-a^{2},-a^{2},-a^{2}),  \notag
\end{equation}%
where $a(t)$ is the scale factor, leads to the equations%
\begin{equation}
\dot{H}=-4\pi G\dot{\phi}^{2},\qquad H^{2}=\frac{4\pi G}{3}\left( \dot{\phi}%
^{2}+2V(\phi )\right) ,\qquad \ddot{\phi}+3H\dot{\phi}+V^{\prime }(\phi )=0,
\label{frie}
\end{equation}%
where $H=\dot{a}/a$ is the Hubble parameter, in both cases (\ref{action})
and (\ref{sqgeq}).

We introduce the coupling\footnote{%
For the purposes of this paper, we can assume $\dot{\phi}>0$, $\alpha >0$.}%
\begin{equation}
\alpha =\frac{\hat{\kappa}\dot{\phi}}{2H}=\sqrt{-\frac{\dot{H}}{3H^{2}}},
\label{alf}
\end{equation}%
where $\hat{\kappa}=\sqrt{16\pi G/3}$. Using $\dot{\phi}$ $=2\alpha H/\hat{%
\kappa}$ inside the second equation (\ref{frie}), we obtain the potential $%
V(\phi (\alpha ))$ as a function of $\alpha $:%
\begin{equation}
V=\frac{2H^{2}}{\hat{\kappa}^{2}}\left( 1-\alpha ^{2}\right) .  \label{Va}
\end{equation}%
Eliminating $\ddot{\phi}$ from the last equation (\ref{frie}), it is easy to
show that $\alpha $ satisfies%
\begin{equation}
\frac{\dot{\alpha}}{H}=-3\alpha (1-\alpha ^{2})-\frac{\hat{\kappa}}{2}\frac{%
V^{\prime }}{H^{2}}.  \label{equa}
\end{equation}%
Introducing the conformal time%
\begin{equation}
\tau =-\int_{t}^{+\infty }\frac{\mathrm{d}t^{\prime }}{a(t^{\prime })},
\label{tau}
\end{equation}%
equation (\ref{equa}) can be converted into the beta function 
\begin{equation}
\beta _{\alpha }\equiv \frac{\mathrm{d}\alpha }{\mathrm{d\ln }|\tau |}=-%
\frac{1}{v}\frac{\dot{\alpha}}{H},  \label{beta}
\end{equation}%
of the cosmic RG flow, where $v\equiv -(aH\tau )^{-1}$.

By differentiating its definition with respect to $\tau $, it is easy to
show that the function $v$ satisfies the linear differential equation%
\begin{equation}
\beta _{\alpha }\frac{\mathrm{d}v}{\mathrm{d}\alpha }=1-3\alpha ^{2}-v,
\label{ev}
\end{equation}%
which can be integrated by quadratures and has solution%
\begin{equation}
v(\alpha )=1-3\alpha ^{2}+6\int_{\alpha _{0}}^{\alpha }\mathrm{d}\alpha
^{\prime }\alpha ^{\prime }\exp \left( -\int_{\alpha ^{\prime }}^{\alpha }%
\frac{\mathrm{d}\alpha ^{\prime \prime }}{\beta _{\alpha }(\alpha ^{\prime
\prime })}\right) ,  \label{va}
\end{equation}%
where $\alpha _{0}$ must be chosen to eliminate the essential singularity.

From the definition of $\alpha $, we have the equation $\dot{H}=-3\alpha
^{2}H^{2}$. If $H$ is viewed as a function of $\alpha $, the equation can be
written as%
\begin{equation*}
\frac{\mathrm{d}H}{\mathrm{d}\alpha }=\frac{3\alpha ^{2}H}{v\beta _{\alpha }}%
,
\end{equation*}%
which is solved by 
\begin{equation}
H(\alpha )=H_{0}\exp \left( \int_{\alpha _{0}}^{\alpha }\frac{3\alpha
^{\prime 2}\mathrm{d}\alpha ^{\prime }}{v(\alpha ^{\prime })\beta _{\alpha
}(\alpha ^{\prime })}\right) ,  \label{ha}
\end{equation}%
where $H_{0}$ is an arbitrary constant.

Dividing $\dot{\phi}$ $=2\alpha H/\hat{\kappa}$ by $\dot{\alpha}=-Hv\beta
_{\alpha }$, we obtain the equation satisfied by $\phi $, also viewed as a
function of $\alpha $, which reads%
\begin{equation}
\frac{\mathrm{d}\phi }{\mathrm{d}\alpha }=\frac{\dot{\phi}}{\dot{\alpha}}=-%
\frac{2\alpha }{\hat{\kappa}v\beta _{\alpha }}  \label{faeq}
\end{equation}%
and is solved by%
\begin{equation}
\phi (\alpha )=-\frac{2}{\hat{\kappa}}\int^{\alpha }\frac{\alpha ^{\prime }%
\mathrm{d}\alpha ^{\prime }}{v(\alpha ^{\prime })\beta _{\alpha }(\alpha
^{\prime })},  \label{fa}
\end{equation}%
where the lower limit of integration remains arbitrary.

By inserting the inverse $\alpha (\phi )$ of this function inside (\ref{Va}%
), we can fully reconstruct the potential $V(\phi )$ from the beta function $%
\beta _{\alpha }$.

\bigskip

In this paper, we assume a generic de-Sitter free beta function%
\begin{equation}
\beta _{\alpha }(\alpha )=\alpha ^{2}(b_{0}+b_{1}\alpha +b_{2}\alpha
^{2}+b_{3}\alpha ^{3}+\cdots )  \label{betaa}
\end{equation}%
that behaves like the one of a quantum field theory around a free-field
fixed point. This means, in practice, that the linear term is missing in (%
\ref{betaa}). In most cases, the first nonvanishing coefficient is negative,
so the de Sitter fixed point $\alpha =0$ corresponds to the infinite past.

In the next section we classify the types of potentials obtained from (\ref%
{betaa}). The other way around (i.e., search for the cosmic RG flow
associated with a given potential)\ is considered in section \ref{pottoflow}%
. Although it is always possible to associate a potential to a flow with
beta function (\ref{betaa}), it is not always possible to generate a beta
function of the form (\ref{betaa}) from a potential, with $\alpha $ defined
as in (\ref{alf}).

The running coupling $\alpha (-\tau )$ is defined from%
\begin{equation*}
\ln \frac{\tau }{\tau ^{\prime }}=\int_{\alpha (-\tau ^{\prime })}^{\alpha
(-\tau )}\frac{\mathrm{d}\alpha ^{\prime }}{\beta _{\alpha }(\alpha ^{\prime
})}.
\end{equation*}%
We often denote $\alpha (-\tau )$ by $\alpha $ and $\alpha (1/k)$ by $\alpha
_{k}$, where $k$ is just a constant for the moment (later on it will denote
the absolute value of the space momentum $\mathbf{k}$ of the quantum
fluctuations of the metric). Writing $\tau ^{\prime }=-1/k$, we have%
\begin{equation*}
\ln (-k\tau )=\int_{\alpha _{k}}^{\alpha }\frac{\mathrm{d}\alpha ^{\prime }}{%
\beta _{\alpha }(\alpha ^{\prime })}.
\end{equation*}

It is possible to show that the spectra $\mathcal{P}$ of the tensor and
scalar fluctuations satisfy RG evolution equations in the superhorizon
limit, with vanishing anomalous dimensions \cite{CMBrunning}. Viewing $%
\mathcal{P}$ as functions of $\tau $ and $\alpha $, the equations read%
\begin{equation}
\frac{\mathrm{d}\mathcal{P}}{\mathrm{d}\ln |\tau |}=\left( \frac{\partial }{%
\partial \ln |\tau |}+\beta _{\alpha }(\alpha )\frac{\partial }{\partial
\alpha }\right) \mathcal{P}=0.  \label{RG}
\end{equation}%
Viewing $\mathcal{P}$ as functions of $\alpha $ and $\alpha _{k}$, the
dependence on $\alpha $ drops out and we have 
\begin{equation}
\mathcal{P}=\mathcal{\tilde{P}}(\alpha _{k}),\qquad \frac{\mathrm{d}\mathcal{%
\tilde{P}}(\alpha _{k})}{\mathrm{d}\ln k}=-\beta _{\alpha }(\alpha _{k})%
\frac{\mathrm{d}\mathcal{\tilde{P}}(\alpha _{k})}{\mathrm{d}\alpha _{k}},
\label{noalfa}
\end{equation}%
which means that the spectra depend on the momentum $k$ only through the
running coupling $\alpha _{k}$. A third option is to view the spectra as
functions of $k/k_{\ast }$ and $\alpha _{\ast }=\alpha (1/k_{\ast })$, where 
$k_{\ast }$ is the pivot scale. Then the equations read%
\begin{equation}
\left( \frac{\partial }{\partial \ln k}+\beta _{\alpha }(\alpha _{\ast })%
\frac{\partial }{\partial \alpha _{\ast }}\right) \mathcal{P}(k/k_{\ast
},\alpha _{\ast })=0.  \label{RGeq}
\end{equation}

Most RG techniques known from particle physics in flat spacetime apply to
the cosmic RG flow. They allow us to work out RG improved perturbation
spectra $\mathcal{P}$, resum the leading and subleading logs, simplify the
computations of tilts and running coefficients and easily push the
calculations to high orders \cite{FakeOnScalar}. In the log expansion the
spectra are expanded in powers of $\alpha _{k}$, with the caveat that a
certain product $\alpha _{k}^{n}\ln (-k\tau )$ is considered of order unity
and treated exactly, where $n$ is a positive integer that depends on the
properties of the beta function.

\section{From the flow to the potential}

\setcounter{equation}{0}\label{flowtopot}

In this section we classify the main types of cosmic RG flows and relate
them to classes of known and less known potentials.

The expansion%
\begin{equation}
v=1-3\alpha ^{2}+6b_{0}\alpha ^{3}+6(b_{1}-3b_{0}^{2})\alpha
^{4}+6(b_{2}-7b_{0}b_{1}+12b_{0}^{3})\alpha ^{5}+\mathcal{O}(\alpha ^{6})
\label{vv}
\end{equation}%
of the function $v=-(aH\tau )^{-1}$ can be obtained straightforwardly from
formula (\ref{va}) to arbitrary orders. The other basic quantities $H$, $V$
and $\phi $ are derived as follows. Once $v$ is known, the Hubble parameter $%
H$ is derived from (\ref{ha}), while the field $\phi (\alpha )$ is obtained
from (\ref{fa}). Then formula (\ref{Va}) gives the potential $V$ as a
function of $\alpha $. By inverting $\phi (\alpha )$ we find $\alpha (\phi )$%
, which then allows us to work out $V$ as a function of $\phi $.

We distinguish various cases, corresponding to different classes of
potentials: 1) $b_{0}\neq 0$; 2) $b_{0}=0$, $b_{1}\neq 0$; 3) $b_{0}=b_{1}=0$%
, $b_{2}\neq 0$; etc.

\subsection[Flows of class I]{Flows of class I: $b_{0}\neq 0$}

If $b_{0}\neq 0$ we find the expansions 
\begin{eqnarray*}
H &=&H_{0}\left[ 1+\frac{3\alpha }{b_{0}}+\frac{3(3-b_{1})\alpha ^{2}}{%
2b_{0}^{2}}+\frac{\alpha ^{3}}{2b_{0}^{3}}%
(9+6b_{0}^{2}-9b_{1}+2b_{1}^{2}-2b_{0}b_{2})+\mathcal{O}(\alpha ^{4})\right]
, \\
\hat{\kappa}\phi &=&-\frac{2}{b_{0}}\ln \frac{\alpha }{\Phi _{0}}+\frac{%
2\alpha b_{1}}{b_{0}^{2}}-\frac{\alpha ^{2}}{b_{0}^{3}}%
(3b_{0}^{2}+b_{1}^{2}-b_{0}b_{2})+\mathcal{O}(\alpha ^{3}), \\
V &=&\frac{2H_{0}^{2}}{\hat{\kappa}^{2}}\left[ 1+\frac{6\alpha }{b_{0}}+%
\frac{\alpha ^{2}}{b_{0}^{2}}\left( 18-3b_{1}-b_{0}^{2}\right) +\frac{%
2\alpha ^{3}}{b_{0}^{3}}(18-9b_{1}+b_{1}^{2}-b_{0}b_{2})+\mathcal{O}(\alpha
^{4})\right] ,
\end{eqnarray*}%
where $H_{0}$ and $\Phi _{0}$ are arbitrary constants. We see that $\Phi
\equiv \Phi _{0}\mathrm{e}^{-b_{0}\hat{\kappa}\phi /2}$ is a power series in 
$\alpha $ that starts from $\mathcal{O}(\alpha )$. Inverting for $\alpha $,
we obtain%
\begin{equation*}
\alpha (\phi )=\Phi +\frac{b_{1}}{b_{0}}\Phi ^{2}-\frac{%
3b_{0}^{2}-2b_{1}^{2}-b_{0}b_{2}}{2b_{0}^{2}}\Phi ^{3}+\mathcal{O}(\Phi
^{4}).
\end{equation*}%
Inserting $\alpha (\phi )$ into $V$, we get the potential%
\begin{eqnarray*}
V &=&\frac{2H_{0}^{2}}{\hat{\kappa}^{2}}\left[ 1+6\frac{\Phi }{b_{0}}%
+(18-b_{0}^{2}+3b_{1})\frac{\Phi ^{2}}{b_{0}^{2}}\right. \\
&&\qquad \left. +(b_{0}b_{2}-b_{0}^{2}(9+2b_{1})+2(18+9b_{1}+b_{1}^{2}))%
\frac{\Phi ^{3}}{b_{0}^{3}}+\mathcal{O}(\Phi ^{4})\right] .
\end{eqnarray*}

The conclusion is that when $b_{0}\neq 0$, the potential is an expansion in
powers of $\Phi $. The de Sitter limit is $\phi \rightarrow \infty \times $%
sign$(b_{0})$, where $V$ tends to a nonvanishing positive constant $%
V_{0}=2H_{0}^{2}/\hat{\kappa}^{2}$, $H_{0}$ being the limiting value of $H$.

The autonomous first order differential equation (\ref{auto}) associated
with this type of potentials was also discussed in \cite{kiritsis} in the
context of holography, but not related to the RG properties of the spectra.

Examples of potentials that fall into this class are the $\alpha $-attractor
E-Models \cite{alfat}, which include the Starobinsky potential%
\begin{equation}
V(\phi )=\frac{m_{\phi }^{2}}{2\hat{\kappa}^{2}}\left( 1-\mathrm{e}^{\hat{%
\kappa}\phi }\right) ^{2},  \label{staropot}
\end{equation}%
the $\alpha $-attractor T-Models \cite{alfat}, which include $V(\phi
)=V_{0}\tanh ^{2}(c\phi )$ and $V(\phi )=V_{0}\tanh ^{4}(c\phi )$ (which can
also be obtained in the context of Palatini inflation \cite{Palatini}), the
potential $V(\phi )=V_{0}(1-\mathrm{e}^{c\phi })$ of exponential SUSY
inflation \cite{expinfl} and the potential $V(\phi )=V_{0}(1-$ sech$(c\phi
)) $ of mutated hilltop inflation \cite{mutatedhilltop}, where $V_{0}$ and $%
c $ are constants.

The running coupling can be worked out from 
\begin{equation}
\mathrm{d}\log {\eta }=\frac{\mathrm{d}\alpha }{\beta _{\alpha }(\alpha )},
\label{diff}
\end{equation}%
where $\eta =-k\tau $. Integrating both sides from $\eta =1$ to generic $%
\eta $, we find 
\begin{equation}
\alpha (-\tau )=\frac{\alpha _{k}}{\lambda },\qquad \lambda \equiv
1-b_{0}\alpha _{k}\log {\eta },  \label{LLb0}
\end{equation}%
to the leading log order and 
\begin{equation}
\alpha (-\tau )=\frac{\alpha _{k}}{\lambda }\prod_{n=1}^{\infty }(1+\alpha
_{k}^{n}\gamma _{n}(\lambda )),  \label{runb0}
\end{equation}%
in general, where 
\begin{equation}
\gamma _{1}(\lambda )=-\frac{b_{1}}{b_{0}}\frac{\ln {\lambda }}{\lambda }%
,\qquad \gamma _{2}(\lambda )=\frac{(\lambda
-1)(b_{1}^{2}-b_{0}b_{2})+b_{1}^{2}(\ln \lambda -1)\ln \lambda }{\lambda
^{2}b_{0}^{2}},  \label{runfb0}
\end{equation}%
etc.

\subsection[Flows of class II]{Flows of class II: $b_{0}=0$, $b_{1}\neq 0$}

If $b_{0}=0$, $b_{1}\neq 0$, we find%
\begin{eqnarray}
H &=&H_{0}\alpha ^{3/b_{1}}\left[ 1-\frac{3b_{2}\alpha }{b_{1}^{2}}+\frac{%
3(3b_{1}^{3}+{3b_{2}^{2}}+b_{1}b_{2}^{2}-b_{1}^{2}b_{3})\alpha ^{2}}{%
2b_{1}^{4}}+\mathcal{O}(\alpha ^{3})\right] ,  \notag \\
\hat{\kappa}\phi &=&\hat{\kappa}\phi _{0}+\frac{2}{b_{1}\alpha }+\frac{2b_{2}%
}{b_{1}^{2}}\ln \alpha -\frac{2\alpha }{b_{1}^{3}}%
(3b_{1}^{2}+b_{2}^{2}-b_{1}b_{3})+\mathcal{O}(\alpha ^{2}).  \label{HII}
\end{eqnarray}%
The de Sitter fixed point is $\phi \times $sign$(b_{1})\rightarrow \infty $.
With no loss of generality, we set the arbitrary constant $\phi _{0}$ to
zero. We see that $\phi (\alpha )$ involves both $\ln \alpha $ and $1/\alpha 
$. We can invert $\alpha (\phi )$ as an expansion in powers of $1/\phi $ and
logarithms of $\phi $. A special case where we can proceed straightforwardly
is $b_{2}=0$, where we find (for $b_{1}\phi >0$)%
\begin{eqnarray}
\alpha &=&\frac{2}{b_{1}\hat{\kappa}\phi }\left[ 1-\frac{4(3b_{1}-b_{3})}{%
b_{1}^{3}\hat{\kappa}^{2}\phi ^{2}}+\frac{4b_{4}}{b_{1}^{4}\hat{\kappa}%
^{3}\phi ^{3}}+\mathcal{O}\left( (\hat{\kappa}\phi )^{-4}\right) \right] , 
\notag \\
V(\phi ) &=&\frac{2H_{0}^{2}}{\hat{\kappa}^{2}}\left( \frac{b_{1}\hat{\kappa}%
\phi }{2}\right) ^{-6/b_{1}}\left[ 1-\frac{4(9b_{1}+b_{1}^{2}-3b_{3})}{%
b_{1}^{4}\hat{\kappa}^{2}\phi ^{2}}+\frac{8b_{4}}{b_{1}^{5}\hat{\kappa}%
^{3}\phi ^{3}}+\mathcal{O}\left( (\hat{\kappa}\phi )^{-4}\right) \right] .
\label{alfaV}
\end{eqnarray}%
We see that this class of potentials is equal to a nontrivial (possibly
fractional) overall power of $\phi $ times a powers series in $1/\phi $.
When $b_{2}\neq 0$ we obtain corrections of the form $(\hat{\kappa}\phi
)^{-n}\ln ^{m}(\hat{\kappa}\phi )$ with $n\geqslant m\geqslant 1$ inside the
square brackets of formulas (\ref{alfaV}).

From (\ref{diff}), we find the running coupling 
\begin{equation}
\alpha (-\tau )=\frac{\alpha _{k}}{\lambda },\qquad \lambda \equiv \sqrt{%
1-2b_{1}\alpha _{k}^{2}\ln {\eta }},  \label{LLb1}
\end{equation}%
to the leading log order, which suggests that the log expansion must be
defined as the expansion in powers of $\alpha _{k}$, with the caveat that $%
\alpha _{k}^{2}\ln {\eta }$ is considered of order unity and resummed
exactly. In general, $\alpha $ is still of the form (\ref{runb0}). For
instance, when $b_{2}=0$ we have 
\begin{eqnarray}
\gamma _{1}(\lambda ) &=&0,\qquad \gamma _{2}(\lambda )=-\frac{b_{3}\ln {%
\lambda }}{\lambda ^{2}b_{1}},\qquad \gamma _{3}(\lambda )=\frac{(1-\lambda
)b_{4}}{\lambda ^{3}b_{1}},  \notag \\
\gamma _{4}(\lambda ) &=&\frac{(\lambda
^{2}-1)(b_{3}^{2}-b_{1}b_{5})+b_{3}^{2}(3\ln \lambda -2)\ln {\lambda }}{%
2\lambda ^{4}b_{1}^{2}},  \label{runfb1}
\end{eqnarray}%
etc.

Examples of potentials of class II are the powerlike potentials $V(\phi
)=V_{0}\phi ^{n}$ of large field inflation \cite{LFI}, where $V_{0}$ is a
constant (see section \ref{pottoflow}). More general polynomial potentials
are included as well, such as $V(\phi )=V_{0}+V_{1}\phi ^{2}$, $V(\phi
)=V_{1}\phi ^{2}+V_{2}\phi ^{4}$ (mixed large field inflation), $V(\phi
)=V_{0}+V_{1}\phi ^{2}+V_{2}\phi ^{4}$ (Hilltop quartic model and double
well inflation \cite{DWI}), MSSM inflation \cite{MSSMI}, etc., where $V_{i}$
are constants.

\subsection[Flows of class III]{Flows of class III: $b_{0}=b_{1}=0$, $%
b_{2}\neq 0$}

\label{classIII}

Under the assumptions $b_{0}=b_{1}=0$, $b_{2}\neq 0$ we find%
\begin{equation*}
\hat{\kappa}\phi =\frac{1}{b_{2}\alpha ^{2}}-\frac{2b_{3}}{b_{2}^{2}\alpha }-%
\frac{2(3b_{2}^{2}+b_{3}^{2}-b_{2}b_{4})}{b_{2}^{3}}\ln \alpha +\mathcal{O}%
(\alpha ).
\end{equation*}%
Again, the inverse function $\phi (\alpha )$ is quite involved, unless we
make some further assumptions, such as $b_{4}=3b_{2}+(b_{3}^{2}/b_{2})$. In
that case, we obtain%
\begin{eqnarray}
H &=&H_{0}\alpha ^{{-}3b_{3}/b_{2}^{2}}\exp \left( -\frac{3}{b_{2}\alpha }%
\right) \left[ 1+\frac{3(3b_{2}^{2}b_{3}+b_{3}^{3}-b_{2}^{2}b_{5})\alpha ^{2}%
}{2b_{2}^{4}}+\mathcal{O}(\alpha ^{3})\right] ,  \notag \\
\alpha &=&\frac{1}{\sqrt{b_{2}\hat{\kappa}\phi }}\left[ 1-\frac{b_{3}}{b_{2}%
\sqrt{b_{2}\hat{\kappa}\phi }}+\frac{b_{3}^{2}}{2b_{2}^{3}\hat{\kappa}\phi }-%
\frac{3b_{2}^{2}b_{3}+b_{3}^{3}-b_{2}^{2}b_{5}}{b_{2}^{3}(b_{2}\hat{\kappa}%
\phi )^{3/2}}+\mathcal{O}\left( (\hat{\kappa}\phi )^{-2}\right) \right] , 
\notag \\
V(\phi ) &=&\frac{2H_{0}^{2}\left( b_{2}\hat{\kappa}\phi \right)
^{3b_{3}/b_{2}^{2}}}{\hat{\kappa}^{2}}\exp \left( {-}\frac{6}{b_{2}}\sqrt{%
b_{2}\hat{\kappa}\phi }{-6}\frac{b_{3}}{b_{2}^{2}}\right) \left[ 1+\frac{%
3b_{3}^{2}}{b_{2}^{3}\sqrt{b_{2}\hat{\kappa}\phi }}+\mathcal{O}\left( (\hat{%
\kappa}\phi )^{-1}\right) \right] .  \label{HIII}
\end{eqnarray}%
Setting $b_{3}=0$ and choosing the coefficients $b_{2n}$, $n>1$,
appropriately, we may obtain an exponential times a power series:%
\begin{equation*}
V(\phi )=\frac{2H_{0}^{2}}{\hat{\kappa}^{2}}\exp \left( -\frac{6}{b_{2}}%
\sqrt{b_{2}\hat{\kappa}\phi }\right) \left[ 1-\frac{b_{2}^{2}-3b_{5}}{%
b_{2}^{3}\hat{\kappa}\phi }+\mathcal{O}\left( (\hat{\kappa}\phi
)^{-2}\right) \times (\text{power series in }(\hat{\kappa}\phi )^{-1})\text{ 
}\right] .
\end{equation*}

The simplest example is $V=V_{0}\mathrm{e}^{\sqrt{c\phi }}$. To our
knowledge, the potentials of class III have not been considered in the
literature so far \cite{encicl}.

The running coupling to the leading log order is%
\begin{equation*}
\alpha (-\tau )=\frac{\alpha _{k}}{(1-3b_{2}\alpha _{k}^{3}\ln {\eta )}^{1/3}%
}
\end{equation*}%
and the subleading orders can be worked out as before.

\section{Spectral RG\ invariance}

\setcounter{equation}{0}\label{spectralRG}

In this section we outline the strategy of the calculations of the
perturbation spectra. We start from a generic Mukhanov-Sasaki action of the
form%
\begin{equation}
S_{\text{MS}}=\frac{1}{2}\int \mathrm{d}\eta \left[ w^{\prime 2}-hw^{2}+%
\frac{2+\sigma }{\eta ^{2}}w^{2}\right] ,  \label{smuck}
\end{equation}%
where the function $w(\eta )$ stands for a generic fluctuation and $h(\alpha
)=1+\mathcal{O}(\alpha )$, $\sigma (\alpha )=\mathcal{O}(\alpha )$ are
generic power series in $\alpha $. The equation of motion is%
\begin{equation}
w^{\prime \prime }+hw-\frac{2+\sigma }{\eta ^{2}}w=0.  \label{muck}
\end{equation}%
The superhorizon limit of its solution and the RG properties of the spectra
can be studied by decomposing $\eta w(\eta )$ as the sum of a power series $%
Q(\ln \eta )$ in $\ln \eta $ plus a power series $W(\eta )$ in $\eta $ and $%
\ln \eta $, such that $W(\eta )\rightarrow 0$ term-by-term for $\eta
\rightarrow 0$:%
\begin{equation}
\eta w=Q(\ln \eta )+W(\eta ).  \label{decompo}
\end{equation}

Inserting (\ref{decompo}) into (\ref{muck}), we find%
\begin{equation*}
\frac{\mathrm{d}^{2}Q}{(\mathrm{d}\ln \eta )^{2}}-3\frac{\mathrm{d}Q}{%
\mathrm{d}\ln \eta }-\sigma Q=-\eta ^{2}\frac{\mathrm{d}^{2}W}{\mathrm{d}%
\eta ^{2}}+2\eta \frac{\mathrm{d}W}{\mathrm{d}\eta }-h\eta ^{2}(W+Q)+\sigma
W.
\end{equation*}%
The right-hand side is negligible in the superhorizon limit, so both sides
of this equations must be separately zero. We end up with the $Q$ equation%
\begin{equation}
\left( \frac{\mathrm{d}}{\mathrm{d}\ln \eta }-3\right) \frac{\mathrm{d}Q}{%
\mathrm{d}\ln \eta }=\sigma Q.  \label{Qeq}
\end{equation}%
The operator in parenthesis must be inverted perturbatively, to eliminate
contributions proportional to $\eta ^{3}$, which do not belong to $Q$. So
doing, we find%
\begin{equation*}
\frac{\mathrm{d}Q}{\mathrm{d}\ln \eta }=-\frac{1}{3}\frac{1}{1-\frac{1}{3}%
\frac{\mathrm{d}}{\mathrm{d}\ln \eta }}\sigma Q
\end{equation*}%
with an arbitrary initial condition $Q(0)$, or, as in ref. \cite{CMBrunning}%
, 
\begin{equation}
\frac{\mathrm{d}Q}{\mathrm{d}\ln \eta }=-\frac{\sigma }{3}Q-\frac{1}{3}%
\sum_{n=1}^{\infty }3^{-n}\frac{\mathrm{d}^{n}(\sigma Q)}{\mathrm{d}\ln
^{n}\eta },  \label{Pw}
\end{equation}%
where the higher-derivative terms on the right-hand side have to be handled
perturbatively.

Choosing a reference scale $k$, we can also view $Q(\ln \eta )$ as a
function $\tilde{Q}(\alpha ,\alpha _{k})$ of $\alpha $ and $\alpha _{k}$,
satisfying%
\begin{equation}
\beta _{\alpha }\frac{\partial \tilde{Q}}{\partial \alpha }=-\frac{\sigma 
\tilde{Q}}{3}-\frac{1}{3}\sum_{n=1}^{\infty }3^{-n}\left( \beta _{\alpha }%
\frac{\partial }{\partial \alpha }\right) ^{n}(\sigma \tilde{Q}).
\label{Pwt}
\end{equation}%
The general solution of this equation can be written in the form%
\begin{equation}
\tilde{Q}(\alpha ,\alpha _{k})=\tilde{Q}(\alpha _{k})\frac{J(\alpha )}{%
J(\alpha _{k})},  \label{Q}
\end{equation}%
where $Q(0)=\tilde{Q}(\alpha _{k})$ is called \textquotedblleft spectral
normalization\textquotedblright .

We divide the calculation of the spectrum in three steps. The first step is
to determine to spectral normalization $\tilde{Q}(\alpha _{k})$ by solving
equation (\ref{muck}) perturbatively in $\alpha _{k}$ with the Bunch-Davies
vacuum condition\footnote{%
When $h\neq 1$ it is necessary to first switch to different variables $%
\tilde{\eta}$ and $\tilde{w}(\tilde{\eta})$ (see section \ref{massren}).}.
To this purpose, we expand the running coupling $\alpha $ in powers of $%
\alpha _{k}$ and write 
\begin{equation}
h(\alpha )=1+\alpha _{k}\sum_{j=0}^{\infty }\tilde{h}_{j}\alpha
_{k}^{j},\qquad \sigma (\alpha )=\alpha _{k}\sum_{j=0}^{\infty }\sigma
_{j}\alpha _{k}^{j},  \label{hs}
\end{equation}%
where $\sigma _{j}$, $\tilde{h}_{j}$ are functions of $\eta $. Expanding the
function $w$ as well,%
\begin{equation}
w(\eta )=w_{0}(\eta )+\sum_{n=1}^{\infty }\alpha _{k}^{n}w_{n}(\eta ),
\label{wetasstaro}
\end{equation}%
equation (\ref{muck}) gives $w_{0}^{\prime \prime }+w_{0}-2(w_{0}/\eta
^{2})=0$ and%
\begin{equation}
w_{n}^{\prime \prime }+w_{n}-2\frac{w_{n}}{\eta ^{2}}=\frac{1}{\eta ^{2}}%
\sum_{j=0}^{n-1}\sigma _{j}w_{n-1-j}-\sum_{j=0}^{n-1}\tilde{h}%
_{j}w_{n-1-j},\qquad n\geqslant 1.  \label{weq}
\end{equation}%
The solution $w$ must agree with (\ref{Q}) in the superhorizon limit by
means of (\ref{decompo}). The comparison between (\ref{Q}) and (\ref%
{wetasstaro}) allows us to derive $Q(0)=\tilde{Q}(\alpha _{k})$. This step
of the calculation can be done for all the classes of cosmic RG flows at
once, because the results are analytic in the coefficients $b_{i}$ of the
beta function.

The second step is to solve the $Q$ equation (\ref{Pwt}) to determine the
function $J(\alpha )$. Note that $J(\alpha )$ is independent of the mass
renormalization $h$ of (\ref{smuck}) and the Bunch-Davies vacuum condition.
This part of the calculation must be done class by class.

The third step is to build the spectra, which we can achieve by taking
advantage of RG invariance. We quantize (\ref{smuck}) by introducing the
operator 
\begin{equation*}
\hat{w}_{\mathbf{k}}(\eta )=w_{\mathbf{k}}(\eta )\hat{a}_{\mathbf{k}}+w_{-%
\mathbf{k}}^{\ast }(\eta )\hat{a}_{-\mathbf{k}}^{\dagger },
\end{equation*}%
where $\hat{a}_{\mathbf{k}}^{\dagger }$ and $\hat{a}_{\mathbf{k}}$ are
creation and annihilation operators satisfying $[\hat{a}_{\mathbf{k}},\hat{a}%
_{\mathbf{k}^{\prime }}^{\dagger }]=(2\pi )^{3}\delta ^{(3)}(\mathbf{k}-%
\mathbf{k}^{\prime })$. We know that the spectra are RG invariant in the
superhorizon limit. However, the $w$ two-point function does not have this
property. The right perturbation is 
\begin{equation}
w_{\text{RG}}(\eta )\equiv \frac{C}{k^{3/2}}\frac{\eta w(\eta )}{J(\alpha )},
\label{wrg}
\end{equation}%
where $C$ is a constant that depends on the model and the type of
perturbation, while the factor $k^{-3/2}$ is introduced to match the known
cases (see sections \ref{acttospec} and \ref{massren}). It is easy to check
that $w_{\text{RG}}$ is indeed RG invariant in the superhorizon limit, where%
\begin{equation}
w^{\text{RG}}\simeq \frac{C}{k^{3/2}}\frac{\tilde{Q}(\alpha ,\alpha _{k})}{%
J(\alpha )}=\frac{C}{k^{3/2}}\frac{\tilde{Q}(\alpha _{k})}{J(\alpha _{k})}.
\label{wrgsup}
\end{equation}

The power spectrum $\mathcal{P}$ of the fluctuations $w^{\text{RG}}$ is
defined by the two-point function%
\begin{equation}
\langle \hat{w}_{\mathbf{k}}^{\text{RG}}(\eta )\hat{w}_{\mathbf{k}^{\prime
}}^{\text{RG}}(\eta )\rangle =(2\pi )^{3}\delta ^{(3)}(\mathbf{k}+\mathbf{k}%
^{\prime })\frac{2\pi ^{2}}{\varsigma k^{3}}\mathcal{P}.  \label{2pt}
\end{equation}%
Here $\varsigma$ is equal to $16$ and $1$ for the tensor and scalar
perturbations, respectively, and takes into account the normalization of the
polarizations and the sum over them. Using (\ref{wrgsup}), we find 
\begin{equation}
\mathcal{P}=\frac{\varsigma k^{3}}{2\pi ^{2}}\left\vert w_{\mathbf{k}}^{%
\text{RG}}\right\vert ^{2}\simeq \varsigma \frac{|C|^{2}}{2\pi ^{2}}%
\left\vert \frac{\tilde{Q}(\alpha _{k})}{J(\alpha _{k})}\right\vert ^{2},
\label{spectrum}
\end{equation}%
which is RG\ invariant in the superhorizon limit, by construction. The tilts 
$n_{\mathcal{P}}$ and the running coefficients can be calculated
straightforwardly by differentiating the spectra:%
\begin{equation}
n_{\mathcal{P}}-\theta =-\beta _{\alpha }(\alpha _{k})\frac{\partial \ln 
\mathcal{P}}{\partial \alpha _{k}},\qquad \frac{\mathrm{d}^{n}n_{\mathcal{P}}%
}{\mathrm{d}\ln k\hspace{0.01in}^{n}}=\left( -\beta _{\alpha }(\alpha _{k})%
\frac{\partial }{\partial \alpha _{k}}\right) ^{n}n_{\mathcal{P}},
\label{tilts}
\end{equation}%
where $\theta =0$ and $\theta =1$ for the tensor and scalar perturbations,
respectively.

Some remarks are in order before proceeding. Typically, the mass
renormalization $h$ is identically one when the Weyl-squared term $C^{2}$ is
absent and the action is (\ref{action}). Instead, $h$ is nontrivial when the
action is (\ref{sqgeq}) and $C^{2}$ is treated by means of the fakeon
prescription and projection. In the presence of a mass renormalization, a
supplementary step is required to impose the Bunch-Davies vacuum condition.
For this reason, we postpone this part of the investigation to section \ref%
{massren} and first derive the spectra in the case $h\equiv 1$.

The leading behavior of $\sigma $ for $\alpha $ small determines the leading
behavior of the spectrum. Inspired by most common scenarios of class I,
which are investigated in section \ref{acttospec}, we distinguish two cases.
In the first case, $\sigma $ starts from order $\alpha ^{2}$,%
\begin{equation}
\sigma _{\text{t}}(\alpha )=\alpha ^{2}\sum_{n=0}^{\infty }s_{n}^{\text{t}%
}\alpha ^{n},  \label{sigmat}
\end{equation}%
while in the second case $\sigma $ starts linearly in $\alpha $,%
\begin{equation}
\sigma _{\text{s}}(\alpha )=\bar{s}\alpha +\alpha ^{2}\sum_{n=0}^{\infty
}s_{n}^{\text{s}}\alpha ^{n}.  \label{sigmaes}
\end{equation}%
We adopt the subscripts and superscripts \textquotedblleft
t\textquotedblright\ and \textquotedblleft s\textquotedblright\ to
distinguish the two cases, because a $\sigma $ like (\ref{sigmat}) is
typical of the tensor perturbations of class I, while a $\sigma $ like (\ref%
{sigmaes}) is typical of the scalar perturbations, also of class I. Apart
from this, the general approach we adopt here makes no real distinction
between tensor fluctuations and scalar fluctuations. Note that in the common
cases of class II\ $\sigma $ is $\mathcal{O}(\alpha ^{2})$ for both tensor
and scalar perturbations (see section \ref{acttospec}).

The first step of the calculation, which is the derivation of $\tilde{Q}%
(\alpha _{k})$, can be performed right away and, as said, is the same for
all classes of RG flows. If we assume (\ref{sigmat}), it is easy to check
that, to the next-to-next-to-leading (NNL) order, the solution of the
equation (\ref{muck}) that satisfies the Bunch-Davies vacuum condition%
\begin{equation}
w(\eta )\simeq \frac{\mathrm{e}^{i\eta }}{\sqrt{2}}\text{\qquad for }\eta
\rightarrow \infty ,  \label{BD}
\end{equation}%
is (\ref{wetasstaro}) with 
\begin{equation}
w_{0}=W_{0},\qquad w_{1}=0,\qquad w_{2}=\frac{s_{0}}{9}W_{2},\qquad
w_{3}=-b_{0}s_{0}\frac{W_{3}}{18}+\left( 8b_{0}s_{0}+3s_{1}\right) \frac{%
W_{2}}{27},  \label{solet}
\end{equation}%
where the recurring functions $W_{i}$ are defined in formula (\ref{Wi}) of
the appendix. The comparison between $w$ and $Q$ in the superhorizon limit,
done by means of (\ref{decompo}), gives 
\begin{eqnarray}
\tilde{Q}_{\text{t}}(\alpha _{k}) &=&\frac{i}{\sqrt{2}}\left[ 1+\frac{s_{0}}{%
3}\alpha _{k}^{2}\left( 2-\tilde{\gamma}_{M}\right) \right.   \notag \\
&&+\left. \frac{\alpha _{k}^{3}}{18}\left( b_{0}s_{0}\left( 32-28\tilde{%
\gamma}_{M}+6\tilde{\gamma}_{M}^{2}+\pi ^{2}\right) +6s_{1}\left( 2-\tilde{%
\gamma}_{M}\right) \right) \right] +\mathcal{O}(\alpha _{k}^{4}),
\label{Qtit}
\end{eqnarray}%
where $\tilde{\gamma}_{M}$ is also defined in the appendix.

If, instead, we assume (\ref{sigmaes}), the solution of (\ref{muck}) to the
NNL order is (\ref{wetasstaro}) with 
\begin{equation}
w_{0}=W_{0},\qquad w_{1}=\frac{\bar{s}}{9}W_{2},\qquad w_{2}=\frac{\bar{s}%
^{2}}{36}W_{4}-\frac{\bar{s}(\bar{s}+3b_{0})}{108}W_{3}+\frac{6s_{0}+8b_{0}%
\bar{s}-\bar{s}^{2}}{54}W_{2},  \label{soles}
\end{equation}%
which gives%
\begin{eqnarray}
\tilde{Q}_{\text{s}}(\alpha _{k}) &=&\frac{i}{\sqrt{2}}\left[ 1+\frac{\bar{s}%
}{3}(2-\tilde{\gamma}_{M})\alpha _{k}+\frac{s_{0}}{3}\left( 2-\tilde{\gamma}%
_{M}\right) \alpha _{k}^{2}\right. +\frac{b_{0}\bar{s}}{36}\left( 32-28%
\tilde{\gamma}_{M}+6\tilde{\gamma}_{M}^{2}+\pi ^{2}\right) \alpha _{k}^{2} 
\notag \\
&&\left. -\frac{\bar{s}^{2}}{108}\left( 8+20\tilde{\gamma}_{M}-6\tilde{\gamma%
}_{M}^{2}-3\pi ^{2}\right) \alpha _{k}^{2}\right] +\mathcal{O}(\alpha
_{k}^{3}).  \label{Qst}
\end{eqnarray}

The second and third steps of the calculation, which are the derivations of
the functions $J(\alpha )$ and the spectra, respectively, are performed in
the next section class by class.

\section{From the flow to the spectra}

\setcounter{equation}{0}\label{specI}

In this section we derive the functions $J(\alpha )$ and the spectra for the
RG\ flows of classes I, II and III with $h\equiv 1$, without referring to
the origin of the flows from specific actions or models. We give enough
details to derive the results to the next-to-next-to-leading log (NNLL)
order. However, once the procedure is clear enough we just report them to
the NLL order.

\subsection{Class I}

If we assume the expansion (\ref{sigmat}), the $Q$ function 
\begin{equation}
\tilde{Q}_{\text{t}}(\alpha ,\alpha _{k})=\tilde{Q}_{\text{t}}(\alpha _{k})%
\frac{J_{\text{t}}(\alpha )}{J_{\text{t}}(\alpha _{k})}  \label{Qtt}
\end{equation}%
is determined by solving (\ref{Pwt}). We obtain 
\begin{equation*}
J_{\text{t}}(\alpha )=1-\frac{s_{0}\alpha }{3b_{0}}+\left(
-2b_{0}^{2}s_{0}+3b_{1}s_{0}+s_{0}^{2}-3b_{0}s_{1}\right) \frac{\alpha ^{2}}{%
18b_{0}^{2}}+\alpha ^{3}\Delta J_{\text{t}}^{\text{I}}+\mathcal{O}(\alpha
^{4}),
\end{equation*}%
where the NNLL corrections $\Delta J_{\text{t}}^{\text{I}}$ are given in
formula (\ref{dIt}). The tensor spectrum is, from (\ref{spectrum}) and (\ref%
{Qtit}),%
\begin{eqnarray}
\mathcal{P}_{\text{t}}^{\text{I}} &=&\frac{4|C_{\text{t}}|^{2}}{\pi ^{2}}%
\left\{ 1+\frac{2s_{0}\alpha _{k}}{3b_{0}}+\frac{\alpha _{k}^{2}}{9b_{0}^{2}}%
\left[ 2b_{0}^{2}\left( 7-3\gamma _{M}\right)
s_{0}+s_{0}(2s_{0}-3b_{1})+3b_{0}s_{1}\right] \right.  \notag \\
&&\qquad \left. +\alpha _{k}^{3}\Delta P_{\text{t}}^{\text{I}}+\mathcal{O}%
(\alpha _{k}^{4})\right\} ,  \label{PtI}
\end{eqnarray}%
where $\Delta P_{\text{t}}^{\text{I}}$ is also given in formula (\ref{dIt}).

If we assume the expansion (\ref{sigmaes}), we find%
\begin{equation}
\tilde{Q}_{\text{s}}(\alpha ,\alpha _{k})=\tilde{Q}_{\text{s}}(\alpha _{k})%
\frac{J_{\text{s}}(\alpha )}{J_{\text{s}}(\alpha _{k})}=\tilde{Q}_{\text{s}%
}(\alpha _{k})\frac{\alpha ^{-\bar{s}/(3b_{0})}}{\alpha _{k}^{-\bar{s}%
/(3b_{0})}}\frac{\tilde{J}_{\text{s}}(\alpha )}{\tilde{J}_{\text{s}}(\alpha
_{k})},  \label{Qts}
\end{equation}%
where 
\begin{equation}
\tilde{J}_{\text{s}}(\alpha )=1+\frac{\alpha }{27b_{0}^{2}}\left( -3b_{0}^{2}%
\bar{s}+9b_{1}\bar{s}+b_{0}\bar{s}^{2}-9b_{0}s_{0}\right) +\alpha ^{2}\Delta 
\tilde{J}_{\text{s}}^{\text{I}}+\mathcal{O}(\alpha ^{3}),  \label{Jts}
\end{equation}%
and $\Delta \tilde{J}_{\text{s}}^{\text{I}}$ is given in formula (\ref{dIs}%
). Using (\ref{Qst}), the spectrum is%
\begin{equation}
\mathcal{P}_{\text{s}}^{\text{I}}=\frac{|C_{\text{s}}|^{2}}{4\pi ^{2}}\alpha
_{k}^{2\bar{s}/(3b_{0})}\left\{ 1-\frac{2\alpha _{k}}{27b_{0}^{2}}\left[
9b_{1}\bar{s}-3b_{0}^{2}\bar{s}(7-3\gamma _{M})+b_{0}(\bar{s}^{2}-9s_{0})%
\right] +\alpha _{k}^{2}\Delta P_{\text{s}}^{\text{I}}+\mathcal{O}(\alpha
_{k}^{3})\right\} .  \label{PsI}
\end{equation}%
The NNLL corrections $\Delta P_{\text{s}}^{\text{I}}$ , which we do not
report explicitly, can be easily derived from (\ref{Qts}), (\ref{Qst}) and (%
\ref{Jts}).

Assuming that the subscripts \textquotedblleft t\textquotedblright\ and
\textquotedblleft s\textquotedblright\ stand for the tensor and scalar
perturbations, respectively, which is what happens in typical cases, we can
define the tensor-to-scalar ratio $r$. The main results to the leading log
order are \ \ \ \ \ \ \ \ \ \ \ \ \ \ \ \ \ \ \ \ \ \ \ \ \ \ \ \ \ \ \ \ \
\ \ \ \ \ \ \ \ \ \ \ \ \ \ \ \ \ \ \ \ \ \ \ \ \ \ \ \ \ \ \ \ \ \ \ \ \ \
\ \ \ \ \ \ \ \ \ \ \ \ \ \ \ \ \ \ \ \ \ \ \ \ \ \ \ \ \ \ \ \ \ \ \ \ \ \
\ \ \ \ \ \ \ \ \ \ \ \ \ \ \ \ \ \ \ \ \ \ \ \ \ \ \ \ \ \ \ 
\begin{eqnarray*}
\mathcal{P}_{\text{t}}^{\text{I}} &\simeq &\frac{4|C_{\text{t}}|^{2}}{\pi
^{2}},\qquad \mathcal{P}_{\text{s}}^{\text{I}}\simeq \frac{|C_{\text{s}}|^{2}%
}{4\pi ^{2}}\alpha _{k}^{2\bar{s}/(3b_{0})},\qquad r^{\text{I}}=\frac{%
\mathcal{P}_{\text{t}}^{\text{I}}}{\mathcal{P}_{\text{s}}^{\text{I}}}\simeq 
\frac{16|C_{\text{t}}|^{2}}{|C_{\text{s}}|^{2}}\alpha _{k}^{-2\bar{s}%
/(3b_{0})}, \\
n_{\text{t}} &=&-\beta _{\alpha }(\alpha _{k})\frac{\partial \ln \mathcal{P}%
_{\text{t}}^{\text{I}}}{\partial \alpha _{k}}\simeq -\frac{2s_{0}^{\text{t}}%
}{3}\alpha _{k}^{2},\qquad \qquad n_{\text{s}}-1=-\beta _{\alpha }(\alpha
_{k})\frac{\partial \ln \mathcal{P}_{\text{s}}^{\text{I}}}{\partial \alpha
_{k}}\simeq -\frac{2\bar{s}}{3}\alpha _{k},
\end{eqnarray*}%
where we have distinguished the coefficients of the two $\sigma $ expansions
by means of superscripts \textquotedblleft t\textquotedblright\ and
\textquotedblleft s\textquotedblright , when necessary.

The novel feature of the general spectra just found is that they can violate
the \textquotedblleft consistency condition\textquotedblright\ $r^{\text{I}%
}+8n_{\text{t}}=0$ to the lowest order. Moreover, the overall factor of $%
\mathcal{P}_{\text{s}}^{\text{I}}$ can be any power of $\alpha _{k}$,
possibly fractional. This is not in contradiction with the present
knowledge. Rather, our approach evades the assumptions of the more common
approaches. Indeed, when we specialize to the slow-roll single-field
inflation driven by the action (\ref{action}), we always find $\bar{s}%
=-3b_{0}$ (so the overall power of $\mathcal{P}_{\text{s}}^{\text{I}}$ is
just $1/\alpha _{k}^{2}$) and the other coefficients conspire to make $r^{%
\text{I}}+8n_{\text{t}}=0$ hold true to the lowest order (see section \ref%
{acttospec}).

Usually, the physical predictions are expressed through the number of
e-foldings $N$. We prefer not to do so, because $N$ is not a perturbative
quantity. Besides, in the RG approach it is more convenient to use
quantities that have clear RG evolution properties. The coupling $\alpha
_{k} $ is not physical (like the coupling constants of quantum field
theory), since we can always make arbitrary perturbative redefinitions%
\begin{equation*}
\alpha _{k}\rightarrow \alpha _{k}+c_{1}\alpha _{k}^{2}+c_{2}\alpha
_{k}^{3}+\cdots ,
\end{equation*}%
without spoiling the basic properties of the flow, where $c_{i}$ are
constants. A way out is to eliminate $\alpha _{k}$ from one physical
quantity (say $|n_{\text{s}}-1|$, which is the best measured one) and then
express every other predictions in terms of it. For example, to the leading
order we find the relation 
\begin{equation}
r^{\text{I}}(1-n_{\text{s}})^{2\bar{s}/(3b_{0})}\simeq \frac{16|C_{\text{t}%
}|^{2}}{|C_{\text{s}}|^{2}}\left( \frac{2\bar{s}}{3}\right) ^{2\bar{s}%
/(3b_{0})}=\text{ constant}.  \label{newrel}
\end{equation}

\subsection{Class II}

\label{specII}

We assume the beta function (\ref{betaa}) with $b_{0}=b_{2}=0$, $b_{1}\neq 0$%
. The expansion (\ref{sigmat}) leads to the $Q$ function 
\begin{equation*}
\tilde{Q}_{\text{t}}(\alpha ,\alpha _{k})=\tilde{Q}_{\text{t}}(\alpha _{k})%
\frac{J_{\text{t}}(\alpha )}{J_{\text{t}}(\alpha _{k})}=\tilde{Q}_{\text{t}%
}(\alpha _{k})\frac{\alpha ^{-s_{0}/(3b_{1})}}{\alpha _{k}^{-s_{0}/(3b_{1})}}%
\frac{\tilde{J}_{\text{t}}(\alpha )}{\tilde{J}_{\text{t}}(\alpha _{k})},
\end{equation*}%
where 
\begin{equation*}
\tilde{J}_{\text{t}}(\alpha )=1-\frac{s_{1}\alpha }{3b_{1}}+\frac{\alpha
^{2}\left(
9b_{3}s_{0}-9b_{1}s_{2}-6b_{1}^{2}s_{0}+b_{1}s_{0}^{2}+3s_{1}^{2}\right) }{%
54b_{1}^{2}}+\alpha ^{3}\Delta \tilde{J}_{\text{t}}^{\text{II}}+\mathcal{O}%
(\alpha ^{4}),
\end{equation*}%
and $\Delta \tilde{J}_{\text{t}}^{\text{II}}$ is given in formula (\ref{dIIt}%
). Then formula (\ref{Qst}) leads to the spectrum%
\begin{eqnarray*}
\mathcal{P}_{\text{t}}^{\text{II}} &=&\frac{4|C_{\text{t}}|^{2}}{\pi ^{2}}%
\alpha _{k}^{2s_{0}/(3b_{1})}\left[ 1+\frac{2s_{1}}{3b_{1}}\alpha _{k}+\frac{%
2s_{0}}{9}(7-3\gamma _{M})\alpha _{k}^{2}\right. \\
&&\left. +\frac{6s_{1}^{2}-b_{1}s_{0}^{2}+9(b_{1}s_{2}-b_{3}s_{0})}{%
27b_{1}^{2}}\alpha _{k}^{2}+\alpha _{k}^{3}\Delta P_{\text{t}}^{\text{II}}+%
\mathcal{O}(\alpha _{k}^{4})\right] ,
\end{eqnarray*}%
where $\Delta P_{\text{t}}^{\text{II}}$ is easy to derive from $\tilde{Q}_{%
\text{t}}(\alpha _{k})$ and $\Delta \tilde{J}_{\text{t}}^{\text{II}}$, when
needed.

From this point onwards we report the results to the NLL order only, since
the procedure is now clear. The expansion (\ref{sigmaes}) gives 
\begin{equation*}
\tilde{Q}_{\text{s}}(\alpha ,\alpha _{k})=\tilde{Q}_{\text{s}}(\alpha _{k})%
\frac{J_{\text{s}}(\alpha )}{J_{\text{s}}(\alpha _{k})}=\tilde{Q}_{\text{s}%
}(\alpha _{k})\frac{\alpha ^{(\bar{s}^{2}-9s_{0})/(27b_{1})}\exp \left( 
\frac{\bar{s}}{3\alpha b_{1}}\right) }{\alpha _{k}{}^{(\bar{s}%
^{2}-9s_{0})/(27b_{1})}\exp \left( \frac{\bar{s}}{3\alpha _{k}b_{1}}\right) }%
\frac{\tilde{J}_{\text{s}}(\alpha )}{\tilde{J}_{\text{s}}(\alpha _{k})},
\end{equation*}%
with%
\begin{equation*}
\tilde{J}_{\text{s}}(\alpha )=1-\frac{81s_{1}+27b_{1}\bar{s}-18s_{0}\bar{s}+2%
\bar{s}^{3}}{243b_{1}}\alpha +\frac{b_{3}\bar{s}}{3b_{1}^{2}}\alpha +%
\mathcal{O}(\alpha ^{2}).
\end{equation*}%
The spectrum is%
\begin{eqnarray*}
\mathcal{P}_{\text{s}}^{\text{II}} &=&\frac{|C_{\text{s}}|^{2}}{4\pi ^{2}}%
\alpha _{k}{}^{2(9s_{0}-\bar{s}^{2})/(27b_{1})}\mathrm{e}^{-2\bar{s}%
/(3\alpha _{k}b_{1})}\left[ 1+\frac{2s_{1}}{3b_{1}}\alpha _{k}+\frac{4\bar{s}%
\alpha _{k}}{243b_{1}}(\bar{s}^{2}-9s_{0})\right. \\
&&\left. +\frac{2\bar{s}}{9}(7-3\gamma _{M})\alpha _{k}-\frac{2b_{3}\bar{s}}{%
3b_{1}^{2}}\alpha _{k}+\mathcal{O}(\alpha _{k}^{2})\right] .
\end{eqnarray*}

The main predictions to the leading order are%
\begin{equation}
r^{\text{II}}=\frac{\mathcal{P}_{\text{t}}^{\text{II}}}{\mathcal{P}_{\text{s}%
}^{\text{II}}}\simeq \frac{16|C_{\text{t}}|^{2}}{|C_{\text{s}}|^{2}}\exp
\left( \frac{2\bar{s}}{3\alpha _{k}b_{1}}\right) ,\qquad n_{\text{t}}\simeq -%
\frac{2s_{0}^{\text{t}}}{3}\alpha _{k}^{2},\qquad n_{\text{s}}-1\simeq -%
\frac{2\bar{s}}{3}\alpha _{k}.  \label{predII}
\end{equation}

Again, we see that the relation $r^{\text{I}}+8n_{\text{t}}=0$ is in general
violated to the lowest order. A new feature here is the presence of
essential singularities in the overall factors of $\mathcal{P}_{\text{s}}^{%
\text{II}}$.

If we assume $\bar{s}=0$, which is what occurs in most known cases (see
sections \ref{acttospec} and \ref{pottoflow}), then $\sigma _{\text{s}}$ has
an expansion of the form (\ref{sigmat}) and we find%
\begin{eqnarray*}
&&r^{\text{II}}\simeq \frac{16|C_{\text{t}}|^{2}}{|C_{\text{s}}|^{2}}\alpha
_{k}{}^{2(s_{0}^{\text{t}}-s_{0}^{\text{s}})/(3b_{1})},\qquad n_{\text{t}%
}\simeq -\frac{2s_{0}^{\text{t}}}{3}\alpha _{k}^{2},\qquad n_{\text{s}%
}-1\simeq -\frac{2s_{0}^{\text{s}}}{3}\alpha _{k}^{2}, \\
&&r^{\text{II}}(1-n_{\text{s}})^{(s_{0}^{\text{s}}-s_{0}^{\text{t}%
})/(3b_{1})}\simeq \frac{16|C_{\text{t}}|^{2}}{|C_{\text{s}}|^{2}}\left( 
\frac{2s_{0}^{\text{s}}}{3}\right) ^{(s_{0}^{\text{s}}-s_{0}^{\text{t}%
})/(3b_{1})}=\text{ constant}.
\end{eqnarray*}

\subsection{Class III}

\label{specIII}

We assume $b_{0}=b_{1}=0$, $b_{2}\neq 0$, $b_{4}=3b_{2}+(b_{3}^{2}/b_{2})$.
We just report the leading order and the overall factors for $b_{3}=0$,
since it is now clear how to compute the corrections, if necessary.

Assuming the expansion (\ref{sigmat}), the $Q$ function is 
\begin{equation*}
\tilde{Q}_{\text{t}}(\alpha ,\alpha _{k})=\tilde{Q}_{\text{t}}(\alpha _{k})%
\frac{\alpha ^{-s_{1}/(3b_{2})}\exp \left( \frac{s_{0}}{3\alpha b_{2}}%
\right) }{\alpha _{k}^{-s_{1}/(3b_{2})}\exp \left( \frac{s_{0}}{3\alpha
_{k}b_{2}}\right) }\frac{\tilde{J}_{\text{t}}(\alpha )}{\tilde{J}_{\text{t}%
}(\alpha _{k})},
\end{equation*}%
where $\tilde{J}_{\text{t}}(\alpha )=1+\mathcal{O}(\alpha )$, so the
spectrum is%
\begin{equation*}
\mathcal{P}_{\text{t}}^{\text{III}}=\frac{4|C_{\text{t}}|^{2}}{\pi ^{2}}%
\alpha _{k}^{2s_{1}/(3b_{2})}\exp \left( -\frac{2s_{0}}{3\alpha _{k}b_{2}}%
\right) \left( 1+\mathcal{O}(\alpha _{k})\right) .
\end{equation*}%
Instead, if we assume (\ref{sigmaes}), we find%
\begin{equation*}
\mathcal{P}_{\text{s}}^{\text{III}}=\frac{|C_{\text{s}}|^{2}}{4\pi ^{2}}%
\alpha _{k}^{2(81s_{1}+2\bar{s}^{3}-18s_{0}\bar{s}-243\bar{s}%
)/(243b_{2})}\exp \left( -\frac{\bar{s}}{3\alpha _{k}^{2}b_{2}}+\frac{2(\bar{%
s}^{2}-9s_{0})}{27\alpha _{k}b_{2}}\right) \left( 1+\mathcal{O}(\alpha
_{k})\right) .
\end{equation*}%
We see that the essential singularity has become more severe.

The main results to the leading order are%
\begin{equation*}
\qquad r^{\text{III}}\simeq \frac{16|C_{\text{t}}|^{2}}{|C_{\text{s}}|^{2}}%
\exp \left( \frac{\bar{s}}{3\alpha _{k}^{2}b_{2}}\right) ,\qquad n_{\text{t}%
}\simeq -\frac{2s_{0}^{\text{t}}}{3}\alpha _{k}^{2},\qquad \qquad n_{\text{s}%
}-1\simeq -\frac{2\bar{s}}{3}\alpha _{k}.
\end{equation*}%
Mimicking the known cases, where typically $\bar{s}=0$ and $s_{0}^{\text{t}%
}=s_{0}^{\text{s}}$, we obtain 
\begin{equation*}
r^{\text{III}}\simeq \frac{16|C_{\text{t}}|^{2}}{|C_{\text{s}}|^{2}}\alpha
_{k}^{2(s_{1}^{\text{t}}-s_{1}^{\text{s}})/(3b_{2})},\qquad n_{\text{t}%
}\simeq -\frac{2s_{0}}{3}\alpha _{k}^{2},\qquad \qquad n_{\text{s}}-1\simeq -%
\frac{2s_{0}}{3}\alpha _{k}^{2}.
\end{equation*}%
In general, the relation $r+8n_{\text{t}}=0$ is violated again.

\section{From the action to the spectra}

\label{acttospec}\setcounter{equation}{0}

In this section we study the RG flows associated with the action (\ref%
{action}). We parametrize the metric as 
\begin{eqnarray}
g_{\mu \nu } &=&\text{diag}(1,-a^{2},-a^{2},-a^{2})+2\text{diag}(\Phi
,a^{2}\Psi ,a^{2}\Psi ,a^{2}\Psi )-\delta _{\mu }^{0}\delta _{\nu
}^{i}\partial _{i}B-\delta _{\mu }^{i}\delta _{\nu }^{0}\partial _{i}B, 
\notag \\
&&-2a^{2}\left( u\delta _{\mu }^{1}\delta _{\nu }^{1}-u\delta _{\mu
}^{2}\delta _{\nu }^{2}+v\delta _{\mu }^{1}\delta _{\nu }^{2}+v\delta _{\mu
}^{2}\delta _{\nu }^{1}\right)  \label{mets}
\end{eqnarray}%
and expand (\ref{action}) to the quadratic order in the fluctuations. These
are the graviton modes $u=u(t,z)$ and $v=v(t,z)$ (chosen to have a space
momentum $\mathbf{k}$ oriented along the $z$ axis after the Fourier
transform) and the scalar mode $\Psi $, while $\Phi $ and $B$ are auxiliary
fields, which can be integrated out straightforwardly. The $\phi $
fluctuation $\delta \phi $ is set to zero by working in the comoving gauge,
where the curvature perturbation $\mathcal{R}$\ coincides with $\Psi $. For
reviews on the parametrizations of the metric fluctuations and their
properties, see \cite{baumann,reviews}.

To study the tensor fluctuations we set $\Phi =\Psi =B=0$. The quadratic
Lagrangian obtained from (\ref{action}) is 
\begin{equation}
(8\pi G)\frac{\mathcal{L}_{\text{t}}}{a^{3}}=\dot{u}_{\mathbf{k}}\dot{u}_{-%
\mathbf{k}}-\frac{k^{2}}{a^{2}}u_{\mathbf{k}}u_{-\mathbf{k}},  \label{lut}
\end{equation}%
plus an identical contribution for $v_{\mathbf{k}}$, where $u_{\mathbf{k}%
}(t) $ is the Fourier transform of $u(t,z)$ with respect to $z$, $\mathbf{k}$
denotes the space momentum and $k=|\mathbf{k}|$. We drop the subscripts $%
\mathbf{k}$ and $-\mathbf{k}$ when no confusion can arise.

To study the scalar fluctuations we set $u=v=0$. Then (\ref{action}) gives
the quadratic Lagrangian 
\begin{equation}
(8\pi G)\frac{\mathcal{L}_{\text{s}}}{a^{3}}=-3(\dot{\Psi}+H\Phi )^{2}+4\pi G%
\dot{\phi}^{2}\Phi ^{2}+\frac{k^{2}}{a^{2}}\left[ 2B(\dot{\Psi}+H\Phi )+\Psi
(\Psi -2\Phi )\right] ,  \notag
\end{equation}%
having Fourier transformed the space coordinates to momentum space and
omitted the subscripts $\mathbf{k}$ and $-\mathbf{k}$. As said, $B$ and $%
\Phi $ appear as auxiliary fields. Once they are integrated out, we obtain $%
\Phi =-\dot{\Psi}/H$ and%
\begin{equation}
(8\pi G)\frac{\mathcal{L}_{\text{s}}}{a^{3}}=3\alpha ^{2}\left( \dot{\Psi}%
^{2}-\frac{k^{2}}{a^{2}}\Psi ^{2}\right) .  \label{ls}
\end{equation}

The Lagrangians (\ref{lut}) and (\ref{ls}) are then converted to the form (%
\ref{smuck}) by defining%
\begin{equation}
w_{\text{t}}=au\sqrt{\frac{k}{4\pi G}},\qquad w_{\text{s}}=\alpha a\Psi 
\sqrt{\frac{3k}{4\pi G}},  \label{w}
\end{equation}%
and switching to the variable $\eta =-k\tau $. We obtain%
\begin{equation}
S_{\text{t,s}}=\frac{1}{2}\int \mathrm{d}\eta \left[ w_{\text{t,s}}^{\prime 
\hspace{0.01in}2}-w_{\text{t,s}}^{2}+(2+\sigma _{\text{t,s}})\frac{w_{\text{%
t,s}}^{2}}{\eta ^{2}}\right] ,  \label{sred}
\end{equation}%
where \cite{CMBrunning}%
\begin{eqnarray}
\sigma _{\text{t}} &=&\frac{2}{v^{2}}\left( 1-\frac{3}{2}\alpha ^{2}\right)
-2,  \notag \\
\sigma _{\text{s}} &=&\left( \frac{\beta _{\alpha }}{\alpha }-\frac{1}{v}%
-1+\beta _{\alpha }\frac{\mathrm{d}}{\mathrm{d}\alpha }\right) \left( \frac{%
\beta _{\alpha }}{\alpha }-\frac{1}{v}\right) -2.  \label{sigma}
\end{eqnarray}%
Using (\ref{betaa}) and (\ref{vv}), we find the expansions%
\begin{eqnarray}
\sigma _{\text{t}} &=&9\alpha ^{2}-24b_{0}\alpha
^{3}+12(3+6b_{0}^{2}-2b_{1})\alpha ^{4}+\mathcal{O}(\alpha ^{5}),  \notag \\
\sigma _{\text{s}} &=&-3b_{0}\alpha +(9+2b_{0}^{2}-3b_{1})\alpha
^{2}-(30b_{0}-5b_{0}b_{1}+3b_{2})\alpha ^{3}+\mathcal{O}(\alpha ^{4}).
\label{sts}
\end{eqnarray}

We know that the fluctuations $w_{\text{RG}}=u$ and $w_{\text{RG}}=\Psi $
are RG invariant in the superhorizon limit. Comparing (\ref{wrg}) with (\ref%
{w}), we find 
\begin{equation}
J_{\text{t}}(\alpha )=\frac{C_{\text{t}}}{vH\sqrt{4\pi G}},\qquad J_{\text{s}%
}(\alpha )=\frac{C_{\text{s}}\alpha }{vH}\sqrt{\frac{3}{4\pi G}}.
\label{ctcs}
\end{equation}
The constants $C_{\text{t}}$ and $C_{\text{s}}$ can be read from the leading
behaviors of these functions around $\alpha \sim 0$. In all classes we find%
\begin{equation}
C_{\text{t}}=H_{0}\sqrt{4\pi G},\qquad C_{\text{s}}=\frac{C_{\text{t}}}{%
\sqrt{3}}.  \label{ctcsI}
\end{equation}%
From (\ref{spectrum}) we derive the spectra%
\begin{equation}
\mathcal{P}_{\text{t}}=\frac{32G}{\pi }v^{2}(\alpha _{k})H^{2}(\alpha
_{k})\left\vert \tilde{Q}_{\text{t}}(\alpha _{k})\right\vert ^{2},\qquad 
\mathcal{P}_{\text{s}}=\frac{2G}{3\pi \alpha _{k}^{2}}v^{2}(\alpha
_{k})H^{2}(\alpha _{k})\left\vert \tilde{Q}_{\text{s}}(\alpha
_{k})\right\vert ^{2}.  \label{PtPs}
\end{equation}%
The tensor-to-scalar ratio is%
\begin{equation*}
r=48\alpha _{k}^{2}\frac{\left\vert \tilde{Q}_{\text{t}}(\alpha
_{k})\right\vert ^{2}}{\left\vert \tilde{Q}_{\text{s}}(\alpha
_{k})\right\vert ^{2}}.
\end{equation*}%
Since $\tilde{Q}(\alpha _{k})=i(1+\mathcal{O}(\alpha _{k}))/\sqrt{2}$ in
every case, the leading contribution to $r$ is always $\simeq 48\alpha
_{k}^{2}$. Assuming $\beta _{\alpha }=b_{n}\alpha ^{n+2}+\mathcal{O}(\alpha
^{n+3})$ to cover all classes of RG flows at the same time, the tilts to the
leading order are 
\begin{equation}
n_{\text{t}}=-\beta _{\alpha }(\alpha _{k})\frac{\partial \ln \mathcal{P}_{%
\text{t}}}{\partial \alpha _{k}}\simeq -6\alpha _{k}^{2},\qquad n_{\text{s}%
}-1=-\beta _{\alpha }(\alpha _{k})\frac{\partial \ln \mathcal{P}_{\text{s}}}{%
\partial \alpha _{k}}\simeq \left\{ 
\begin{tabular}{l}
$2b_{0}\alpha _{k}$ for $n=0$ \\ 
$2(b_{1}-3)\alpha _{k}^{2}$ for $n=1$ \\ 
$-6\alpha _{k}^{2}$ for $n>1$%
\end{tabular}%
\right.  \label{predd}
\end{equation}%
and the relation $r+8n_{\text{t}}\simeq 0$ is always satisfied to the same
order. Since $n_{\text{s}}-1\sim -0.035$ at the pivot scale $k=0.05$Mpc$%
^{-1} $ \cite{Planck18}, we have $r\simeq 0.3$ for $n>1$, which is ruled out
by data \cite{Planck18}. Thus, all the models described by the action (\ref%
{action}) and belonging to classes greater than or equal to III are ruled
out. The models of class II must have $b_{1}<-5.4$ to guarantee $r<0.1$,
while the models of class I must have $b_{0}<-0.38$.

The investigation of the flows for the action (\ref{sqgeq}) is postponed to
section \ref{massren}, because the presence of $C^{2}$ typically turns on a
nontrivial mass renormalization $\Delta h$.

\section{From the potential to the flow}

\setcounter{equation}{0}\label{pottoflow}

Formula (\ref{beta}) does not give a beta function of the form (\ref{betaa})
for any potential $V(\phi )$, with the coupling $\alpha $ defined in (\ref%
{alf}). Here we consider two further models of classes II and III and list
the most important potentials that are studied in the literature, but are
not described by our present approach.

\subsection[Powers]{Powerlike potentials $V(\phi )=\phi ^{2n}$}

The simplest representatives of class II are the powerlike potentials%
\begin{equation}
V=\frac{m^{4-2n}}{(2n)!}\phi ^{2n},  \label{Vn}
\end{equation}%
where $m$ is a constant of dimension one. We find%
\begin{eqnarray*}
H &=&\frac{m(\hat{\kappa}m)^{1-n}\left( 2n\right) ^{n}}{(3\alpha )^{n}\sqrt{%
2(2n)!}}\left[ 1-\frac{\alpha ^{2}}{2}+\frac{12-5n}{8n}\alpha ^{4}+\mathcal{O%
}(\alpha ^{6})\right] , \\
v &=&1-3\alpha ^{2}-\frac{18}{n}\alpha ^{4}-\frac{18}{n^{2}}(3n+11)\alpha
^{6}+\mathcal{O}(\alpha ^{8}), \\
\hat{\kappa}\phi (\alpha ) &=&c_{0}-\frac{2n}{3\alpha }+\frac{2\alpha }{3}+%
\frac{2}{3n}(n-2)\alpha ^{3}+\mathcal{O}(\alpha ^{4}),
\end{eqnarray*}%
where $c_{0}$ is a constant. The beta function reads%
\begin{equation*}
\beta _{\alpha }=-\frac{3}{n}\alpha ^{3}-\frac{3}{n^{2}}(3n-1)\alpha ^{5}-%
\frac{3}{n^{3}}(9n^{2}+12n+7)\alpha ^{7}+\mathcal{O}(\alpha ^{9}).
\end{equation*}

The results of subsection \ref{specII} give the predictions (\ref{predII}),
to the leading order, at least for generic $h$ and $\sigma $ in (\ref{smuck}%
). With suitable values of $b_{1}$ and $\bar{s}\neq 0$ it is not difficult
to fit the data available at present. However, in the standard cases of
section \ref{acttospec}, formulas (\ref{sts}) give $\bar{s}=0$, so the main
predictions are actually (\ref{predd}) to the leading order, i.e. 
\begin{equation*}
r\simeq 48\alpha _{k}{}^{2},\qquad n_{\text{t}}\simeq -6\alpha
_{k}^{2},\qquad n_{\text{s}}-1\simeq -\frac{6}{n}\left( n+1\right) \alpha
_{k}^{2},\qquad \frac{r}{1-n_{\text{s}}}\simeq \frac{8n}{n+1},
\end{equation*}%
which are ruled out because $r$ is too large. We do not know at present how
to obtain $\bar{s}\neq 0$ from conventional Lagrangian models.

\subsection[New potential]{Class III potential $V(\phi )=\exp \left( \sqrt{-%
\hat{\kappa}\phi }\right) $}

The simplest example of potential of class III is 
\begin{equation}
V(\phi )=V_{0}\exp \left( c\sqrt{-\hat{\kappa}\phi }\right) ,
\label{classIIIpot}
\end{equation}%
where $V_{0}$ and $c$ are positive constants. It can be obtained from the
formulas of subsection \ref{classIII} with%
\begin{equation*}
b_{0}=b_{1}=b_{3}=0,\qquad b_{2}=-\frac{36}{c^{2}},\qquad
b_{4}=3b_{2},\qquad b_{5}=\frac{b_{2}^{2}}{3},\qquad V_{0}=\frac{2H_{0}^{2}}{%
\hat{\kappa}^{2}},
\end{equation*}%
etc. We find%
\begin{equation*}
H=H_{0}\exp \left( \frac{c^{2}}{12\alpha }\right) \left( 1+\mathcal{O}%
(\alpha ^{2})\right) ,\qquad \hat{\kappa}\phi =-\frac{c^{2}}{36\alpha ^{2}}+%
\mathcal{O}(\alpha ).
\end{equation*}%
In the standard cases, where formula (\ref{sts}) gives 
\begin{equation}
\sigma _{\text{t}}=9\alpha ^{2}+36\alpha ^{4}+\mathcal{O}(\alpha
^{5}),\qquad \sigma _{\text{s}}=9\alpha ^{2}+\frac{108}{c^{2}}\alpha ^{3}+%
\mathcal{O}(\alpha ^{4}),  \notag
\end{equation}%
the main predictions to the leading log order are, using (\ref{ctcsI}),%
\begin{eqnarray*}
\mathcal{P}_{\text{t}} &\simeq &\frac{16GH_{0}^{2}}{\pi }\exp \left( \frac{%
c^{2}}{6\alpha _{k}}\right) ,\qquad \mathcal{P}_{\text{s}}\simeq \frac{%
GH_{0}^{2}}{3\pi \alpha _{k}^{2}}\exp \left( \frac{c^{2}}{6\alpha _{k}}%
\right) ,\qquad r\simeq 48\alpha _{k}^{2}, \\
n_{\text{t}} &\simeq &-6\alpha _{k}^{2},\qquad \qquad n_{\text{s}}-1\simeq
-6\alpha _{k}^{2},\qquad \frac{r}{1-n_{\text{s}}}\simeq 8.
\end{eqnarray*}%
Note the essential singularity in both spectra. The model is ruled out by
present data, since it predicts $r\sim 8(1-n_{\text{s}})\sim 0.28$, which is
too large \cite{Planck18}.

In general, it seems that raising the class raises the prediction for $r$
and makes it more difficult to fit the data, unless we find ways to obtain $%
\bar{s}\neq 0$.

\subsection{Potentials not described by the present approach}

It is worth to emphasize that many potentials studied in the literature
cannot be described straightforwardly by the approach pursued so far. This
may mean that they do not have an interpretation in terms of a perturbative
RG flow or simply that they need an upgraded approach, which may, for
example, make use of a more sophisticated definition of coupling $\alpha $.
Examples of such cases are the potentials obtained by including the
radiative corrections proportional to $\ln (\hat{\kappa}\phi )$. We mention
the radiatively corrected massive inflation, the radiatively corrected
quartic inflation, the Coleman-Weinberg inflation and the radiatively
corrected Higgs inflation \cite{radi,linde}. We recall that, instead, powers
of $\ln (\hat{\kappa}\phi )/\phi $ can be obtained in class II when $%
b_{2}\neq 0$.

Other important cases that are left out are (for a more complete list, see 
\cite{encicl}):

-- the potential $V(\phi )=V_{0}[1+\alpha \ln (\hat{\kappa}\phi )]$ of loop
inflation \cite{loop};

-- the potential $V=V_{0}\left[ 1+\cos (\hat{\kappa}\phi )\right] $ of
natural inflation \cite{nat};

-- the potential $V(\phi )=V_{0}\mathrm{e}^{-c\phi }$ of power law inflation 
\cite{PLI};

-- the potential $V(\phi )=V_{0}(\hat{\kappa}\phi )^{2}/[1+c(\hat{\kappa}%
\phi )^{2}]$ of radion gauge inflation \cite{RGI}.

It is also unclear how to describe the potentials associated with generic $%
f(R)$ theories, since they do not easily fall into our classification. For
example, $f(R)=R+cR^{3}$ is equivalent to the potential $V(\phi )=V_{0}%
\mathrm{e}^{\hat{\kappa}\phi /2}(1-\mathrm{e}^{\hat{\kappa}\phi })^{3/2}$ 
\cite{defelice}.

\section{Nontrivial mass renormalization}

\setcounter{equation}{0}\label{massren}

In this section we study the Mukhanov-Sasaki action (\ref{smuck}) in the
presence of a nontrivial mass renormalization%
\begin{equation}
h(\alpha )=1+\bar{h}\alpha +\alpha ^{2}\sum_{n=0}^{\infty }h_{n}\alpha ^{n},
\label{halfa}
\end{equation}%
where $\bar{h}$ and $h_{n}$ are constants. The function $J(\alpha )$ of
equation (\ref{Q}) remains the same, while $\tilde{Q}(\alpha _{k})$ changes.

Following \cite{CMBrunning}, we rewrite the action (\ref{smuck}) as%
\begin{equation}
\tilde{S}_{\text{t}}^{\text{prj}}=\frac{1}{2}\int \mathrm{d}\tilde{\eta}%
\left( \tilde{w}^{\prime \hspace{0.01in}2}-\tilde{w}^{2}+(2+\tilde{\sigma})%
\frac{\tilde{w}^{2}}{\tilde{\eta}^{2}}\right) ,  \label{s2}
\end{equation}%
where the new variable $\tilde{\eta}(\eta )$ is defined as the solution of
the differential equation $\tilde{\eta}^{\prime }(\eta )=\sqrt{h(\eta )}$
with the initial condition $\tilde{\eta}(0)=0$. We have%
\begin{equation}
\tilde{w}(\tilde{\eta}(\eta ))=h(\eta )^{1/4}w(\eta ),\qquad \tilde{\sigma}=%
\frac{\tilde{\eta}^{2}(\sigma +2)}{\eta ^{2}h}+\frac{\tilde{\eta}^{2}}{%
16h^{3}}\left( 4hh^{\prime \prime }-5h^{\prime \hspace{0.01in}2}\right) -2.
\label{wtst}
\end{equation}

Using (\ref{halfa}) we find%
\begin{eqnarray}
\tilde{\eta}(\eta ) &=&\eta \left[ 1+\frac{\bar{h}}{2}\alpha +\frac{\alpha
^{2}}{8}(4h_{0}-4b_{0}\bar{h}-\bar{h}^{2})+b_{0}^{2}\bar{h}\alpha ^{3}\right.
\notag \\
&&\left. +\frac{\alpha ^{3}}{2}(h_{1}-b_{1}\bar{h})+\frac{\alpha ^{3}}{16}(%
\bar{h}+4b_{0})(\bar{h}^{2}-4h_{0})+\mathcal{O}(\alpha ^{4})\right] .
\label{etat}
\end{eqnarray}%
The advantage of (\ref{s2}) is that in the new variables the\ Bunch-Davies
vacuum condition is the usual one,%
\begin{equation}
\tilde{w}(\tilde{\eta})\simeq \frac{\mathrm{e}^{i\tilde{\eta}}}{\sqrt{2}}%
\text{\qquad for }\tilde{\eta}\rightarrow \infty ,  \label{bdt}
\end{equation}%
while it is not so for (\ref{smuck}). We first work out the solution 
\begin{equation*}
\tilde{w}(\tilde{\eta})=\tilde{w}_{0}(\tilde{\eta})+\sum_{n=1}^{\infty
}\alpha _{k}^{n}\tilde{w}_{n}(\tilde{\eta})
\end{equation*}%
of the problem in the new variables, which looks similar to the solutions (%
\ref{solet}) and (\ref{soles}). Then we switch back to the original
variables by means of (\ref{etat}), to derive $w(\eta )$. From $w(\eta )$ we
can read $\tilde{Q}(\alpha _{k})$ by means of (\ref{decompo}) and (\ref{Q}).
Finally, we obtain the spectrum from formula (\ref{spectrum}).

If we assume the tensorial sigma (\ref{sigmat}), the solutions to the NNL
order are 
\begin{eqnarray}
\tilde{w}_{0}(\tilde{\eta}) &=&W_{0}(\tilde{\eta}),\qquad \tilde{w}_{1}(%
\tilde{\eta})=0,\qquad \tilde{w}_{2}(\tilde{\eta})=\frac{s_{0}^{\prime }}{9}%
W_{2}(\tilde{\eta}),  \notag \\
\tilde{w}_{3}(\tilde{\eta}) &=&-b_{0}s_{0}^{\prime }\frac{W_{3}(\tilde{\eta})%
}{18}+\left( 8b_{0}s_{0}^{\prime }+3s_{1}^{\prime }\right) \frac{W_{2}(%
\tilde{\eta})}{27},  \notag
\end{eqnarray}%
where 
\begin{equation*}
s_{0}^{\prime }=s_{0}-\frac{9}{4}b_{0}\bar{h},\qquad s_{1}^{\prime }=s_{1}+%
\frac{9}{4}\left[ 2b_{0}^{2}\bar{h}-b_{1}\bar{h}+b_{0}\left( \bar{h}%
^{2}-2h_{0}\right) \right] .
\end{equation*}

If we assume the scalar sigma (\ref{sigmaes}), the solutions are 
\begin{eqnarray*}
\tilde{w}_{0}(\tilde{\eta}) &=&W_{0}(\tilde{\eta}),\qquad \tilde{w}_{1}(%
\tilde{\eta})=\frac{\bar{s}}{9}W_{2}(\tilde{\eta}), \\
\tilde{w}_{2}(\tilde{\eta}) &=&\frac{\bar{s}^{2}}{36}W_{4}(\tilde{\eta})-%
\frac{\bar{s}(\bar{s}+3b_{0})}{108}W_{3}(\tilde{\eta})+\frac{12s_{0}-27b_{0}%
\bar{h}+16b_{0}\bar{s}-2\bar{s}^{2}}{108}W_{2}(\tilde{\eta}).
\end{eqnarray*}

As before, these results do not depend on the class of RG flow, as is
apparent from the absence of singularities when some coefficients of the
beta function vanish. When we want to calculate the spectra, instead, we
must work class by class. We report the corrections to the spectra to the
same orders as in section \ref{specI}. However, since the NNLL\
contributions are rather lengthy and in most practical applications $\bar{h}$
vanishes, we report those just for $\bar{h}\neq 0$. In class I we have%
\begin{eqnarray}
\mathcal{P}_{\text{t}}^{\text{I,h}} &=&\mathcal{P}_{\text{t}}^{\text{I}}+%
\frac{4|C_{\text{t}}|^{2}}{\pi ^{2}}\left[ -\frac{3}{2}\bar{h}\alpha _{k}+%
\frac{\bar{h}}{8b_{0}}(15\bar{h}b_{0}-4b_{0}^{2}(4-3\gamma
_{M})-8s_{0})\alpha _{k}^{2}\right.  \notag \\
&&\left. -\frac{3}{2}h_{1}\alpha _{k}^{3}+\mathcal{O}(\bar{h})\mathcal{O}%
(\alpha _{k}^{3})+\mathcal{O}(\alpha _{k}^{4})\right] ,  \notag \\
\mathcal{P}_{\text{s}}^{\text{I,h}} &=&\mathcal{P}_{\text{s}}^{\text{I}}+%
\frac{|C_{\text{s}}|^{2}}{4\pi ^{2}}\alpha _{k}^{2\bar{s}/(3b_{0})}\left[ -%
\frac{3}{2}\bar{h}\alpha _{k}+\mathcal{O}(\bar{h})\mathcal{O}(\alpha
_{k}^{2})+\mathcal{O}(\alpha _{k}^{3})\right] .  \label{PIh}
\end{eqnarray}%
The spectra of class II are%
\begin{eqnarray}
\mathcal{P}_{\text{t}}^{\text{II,h}} &=&\mathcal{P}_{\text{t}}^{\text{II}}+%
\frac{4|C_{\text{t}}|^{2}}{\pi ^{2}}\alpha _{k}^{2s_{0}/(3b_{1})}\left[ -%
\frac{3}{2}\bar{h}\alpha _{k}+\frac{1}{8b_{1}}(15b_{1}\bar{h}%
^{2}-12b_{1}h_{0}-8\bar{h}s_{1})\alpha _{k}^{2}\right.  \notag \\
&&\left. -\left( \frac{3}{2}h_{1}+\frac{h_{0}s_{1}}{b_{1}}\right) \alpha
_{k}^{3}+\mathcal{O}(\bar{h})\mathcal{O}(\alpha _{k}^{3})+\mathcal{O}(\alpha
_{k}^{4})\right] ,  \notag \\
\mathcal{P}_{\text{s}}^{\text{II,h}} &=&\mathcal{P}_{\text{s}}^{\text{II}}+%
\frac{|C_{\text{s}}|^{2}}{4\pi ^{2}}\alpha _{k}{}^{2(9s_{0}-\bar{s}%
^{2})/(27b_{1})}\mathrm{e}^{-2\bar{s}/(3\alpha _{k}b_{1})}\left[ -\frac{3}{2}%
\bar{h}\alpha _{k}+\mathcal{O}(\alpha _{k}^{2})\right] .  \label{ppp}
\end{eqnarray}%
The spectra of class III are unmodified to the order they were reported in
section \ref{specI}.

The main applications of these results concern the flows associated with the
action (\ref{sqgeq}). This allows us to generalize the NNLL results found in 
\cite{CMBrunning,FakeOnScalar,ABP} from the Starobinsky potential (\ref%
{staropot}) to a generic flow of class I.

We do not repeat the derivations of \cite{ABP} here, but just recall the
main steps, focusing on the tensor fluctuations first. We expand the action (%
\ref{sqgeq}) by means of (\ref{mets}) to the quadratic order in $u$ and $v$,
setting the scalar fields $\Phi =\Psi =B$ to zero (this is allowed, since
the scalar and tensor perturbations do not mix at the quadratic level). We
obtain a higher-derivative Lagrangian for $u$ plus an identical copy for $v$%
. Then we eliminate the higher derivatives by introducing auxiliary fields
and diagonalize the Lagrangian in the de Sitter limit by means of a change
of variables. We get a new Lagrangian $\mathcal{L}_{\text{t}}(U_{1},U_{2})$
that contains a physical field $U_{1}$ and a fakeon $U_{2}$, is diagonal at $%
\alpha =0$ and, as said, is free of higher derivatives. The fakeon
projection allows us to eliminate $U_{2}$, but, in general, leaves a
nonlocal action. We discover that to the NNLL\ order included the
nonlocalities drop out. This gives a two-derivative action similar to (\ref%
{lut}), to the said order, but not quite the same. The main differences are
the presence of a nonvanishing $\Delta h$ in (\ref{sm}) ($h\neq 1$ in the
notation of (\ref{smuck})), and, crucially, the ABP bound \cite{ABP}, which
tells us when the fakeon projection is valid. We should emphasize that the
methods just recalled are not applicable to the analysis of the corrections
beyond the NNLL order, where the fakeon projection is truly nonlocal. Their
investigation remains an open problem.

We report the results to the NLL order. As in \cite%
{CMBrunning,FakeOnScalar,ABP}, the fakeon projection sets $U_{2}=\mathcal{O}%
(\alpha ^{2})$, so the $w$ action is just given by $\mathcal{L}_{\text{t}%
}(U_{1},0)$ to the NLL order. Working it out, we obtain (\ref{smuck})\ with 
\begin{eqnarray}
h_{\text{t}} &=&1-3\xi \zeta ^{2}\alpha ^{2}+\frac{3\xi \zeta ^{3}\alpha ^{3}%
}{2b_{0}}(6\xi -12+2\xi b_{0}^{2}+b_{0}^{2}\xi ^{2})+\mathcal{O}(\alpha
^{4}),  \notag \\
\sigma _{\text{t}} &=&9\zeta \alpha ^{2}-\frac{3\zeta ^{2}\alpha ^{3}}{2b_{0}%
}(16b_{0}^{2}+18\xi +17\xi b_{0}^{2})+\mathcal{O}(\alpha ^{4}),  \label{hstf}
\end{eqnarray}%
where%
\begin{equation*}
\xi =\frac{4H_{0}^{2}}{m_{\chi }^{2}},\qquad \zeta =\left( 1+\frac{\xi }{2}%
\right) ^{-1},\qquad H_{0}=\sqrt{\frac{8\pi GV_{0}}{3}},
\end{equation*}%
and $V_{0}$ is the value of the potential $V(\phi )$ in the de Sitter limit.
The constant $C_{\text{t}}$ $=H_{0}\sqrt{4\zeta \pi G}$ of (\ref{wrg}) can
be computed in such a limit, where $U_{1}$ and $U_{2}$ decouple and $U_{2}$
can be dropped. At this point, formulas (\ref{hstf}) can be used to work out
the spectrum $\mathcal{P}_{\text{t}}^{\text{I,h}}$ given in (\ref{PIh}).

Although the fakeon $U_{2}$ and its projection participate in the relation
between $u$ and $w$, RG invariance allows us to get directly to the final
result. Indeed, formula (\ref{wrg}) gives $w_{\text{RG}}$, that is to say $u$%
, and allows us to build the RG invariant spectrum directly from $w$. Since
the fakeon projection is RG invariant \cite{CMBrunning}, the result
coincides with the one found by performing the projection explicitly.

It is easy to show that the ABP\ bound $m_{\chi }>m_{\phi }/4$ of ref. \cite%
{ABP}, which is obtained by requiring that the fakeon projection be
consistent throughout the RG\ flow, now becomes $m_{\chi }>H_{0}/2$.

In the case of the scalar perturbations, we find the same ABP\ bound,
together with%
\begin{equation}
h_{\text{s}}=1,\qquad \sigma _{\text{s}}=-3b_{0}\alpha
+(9+2b_{0}^{2}-3b_{1})\alpha ^{2}+\mathcal{O}(\alpha ^{3})  \label{hssf}
\end{equation}%
and $C_{\text{s}}=H_{0}\sqrt{4\pi G/3}$, so $\mathcal{P}_{\text{s}}^{\text{%
I,h}}$ coincides with $\mathcal{P}_{\text{s}}^{\text{I}}$ at this level.

The tensor-to-scalar ratio to the leading order is 
\begin{equation*}
r\simeq 48\zeta \alpha _{k}^{2}.
\end{equation*}%
This means that, due to the ABP bound $m_{\chi }>H_{0}/2$, all the flows of
class I predict $r$ in the interval $0.4\lesssim 1000r\lesssim 3$ \cite{ABP}%
. Nevertheless, the Starobinsky potential (\ref{staropot}) remains special,
since it is the only one that makes the action (\ref{sqgeq}) renormalizable.

The NNLL corrections can be worked out with the methods of \cite%
{FakeOnScalar} for the tensor and scalar perturbations of class I. In
classes II\ and III the nontrivial $H$ expansions of formulas (\ref{HII})
and (\ref{HIII}) pose new challenges and the strategy just described needs
to be upgraded with the help of further redefinitions.

The window $0.4\lesssim 1000r\lesssim 3$ should be partially covered by
near-future experiments, such as LiteBIRD, which is expected to reach an
uncertainty $\delta r<0.001$ \cite{Litebird}. Hopefully, the rest of the
window will be tested in next decades.

We conclude with a comment on the fakeon prescription for the action (\ref%
{sqgeq}). If the degrees of freedom associated with the higher derivatives
are quantized as usual, that is to say, by means of the Feynman
prescription, some of them are ghosts, which makes the theory physically
unacceptable. At first, it may seem that, since the ghosts are massive, and
heavy, they leave no remnants in the superhorizon limit and we should obtain
the same results with ghosts as with fakeons. This is not true, because the
physical predictions depend on the subhorizon limit as well, both because of
the Bunch-Davies condition and because we need to ensure that the\ fakeons
are fake at all scales, including the subhorizon ones. The consequence of
this requirement is the ABP bound mentioned above, which impacts the
physical predictions and has no counterpart in the approaches with ghosts.
For the predictions of the theories with ghosts, see \cite{salvio}. For
detailed comparisons with the predictions of the theories with fakeons, see 
\cite{ABP}.

\section{Conclusions}

\label{conclusions}\setcounter{equation}{0}

We have studied several aspects of the correspondence between the RG flow
familiar from perturbative quantum field theory and the cosmic RG flow,
which provides an alternative approach to inflationary cosmology. The RG
techniques allow us to calculate RG improved perturbation spectra to high
orders. Moreover, RG invariance can be used as a guiding principle to build
the spectra from a general Mukhanov-Sasaki action, in a more axiomatic
spirit, that is to say, without referring to specific models. The resulting
spectra are expansions in powers of the running coupling times certain
prefactors, which may carry essential singularities. They can also evade the
assumptions that imply the relation $r+8n_{\text{t}}=0$ to the leading
order. The classification emerging from the RG\ analysis helps identify the
classes of models that have more chances to fit the data. Not all classes of
potentials considered in the literature can be described efficiently by
means of the cosmic RG approach. Those left out require further
investigations or do not have a beta function of the form considered here.

The RG approach also applies to models that contain purely virtual
particles, such as the theory of quantum gravity $R+R^{2}+C^{2}$ \cite%
{LWgrav}, as well as more general models with actions (\ref{sqgeq}) and
potentials $V(\phi )$. In class I, for example, we find that although the
ABP bound required by the consistency of the formulation depends on the
model, the final prediction for $r$ is universal to the leading order.

\vskip10truept \noindent {\large \textbf{Acknowledgments}}

\vskip 1truept

We thank A. Karam and A. Racioppi for helpful discussions. M.P. is supported
by the Estonian Research Council grants PRG803 and MOBTT86 and by the EU
through the European Regional Development Fund CoE program TK133
\textquotedblleft The Dark Side of the Universe\textquotedblright .

\renewcommand{\thesection}{\Alph{section}} \renewcommand{\theequation}{%
\thesection.\arabic{equation}} \setcounter{section}{0}

\section{Appendix. Reference formulas}

\label{formulas} \setcounter{equation}{0}

In this appendix we collect some formulas used in the paper, starting from
the recurring functions \cite{FakeOnScalar}%
\begin{eqnarray}
W_{0} &=&\frac{i(1-i\eta )}{\eta \sqrt{2}}\mathrm{e}^{i\eta },\qquad W_{2}=3%
\left[ \text{Ei}(2i\eta )-i\pi \right] W_{0}^{\ast }+\frac{6W_{0}}{(1-i\eta )%
},  \notag \\
W_{3} &=&\left[ 6(\ln \eta +\tilde{\gamma}_{M})^{2}+24i\eta
F_{2,2,2}^{1,1,1}\left( 2i\eta \right) +\pi ^{2}\right] W_{0}^{\ast }+\frac{%
24W_{0}}{(1-i\eta )}-4(\ln \eta +1)W_{2},  \label{Wi} \\
W_{4} &=&-\frac{16W_{0}}{1+\eta ^{2}}+\frac{2(13+i\eta )W_{2}}{9(1+i\eta )}+%
\frac{W_{3}}{3}+4G_{2,3}^{3,1}\left( -2i\eta \left\vert _{0,0,0}^{\
0,1}\right. \right) W_{0},  \notag
\end{eqnarray}%
where Ei denotes the exponential-integral function, $F_{b_{1},\cdots
,b_{q}}^{a_{1},\cdots ,a_{p}}(z)$ is the generalized hypergeometric function 
$_{p}F_{q}(\{a_{1},\cdots ,a_{p}\},\{b_{1},\cdots ,b_{q}\};z)$ and $%
G_{p,q}^{m,n}$ is the Meijer G function. Moreover, $\tilde{\gamma}%
_{M}=\gamma _{M}-(i\pi /2)$, $\gamma _{M}=\gamma _{E}+\ln 2$, $\gamma _{E}$
being the Euler-Mascheroni constant.

We also give the NNLL\ contributions to the $J$ functions and the spectra.
For class I we find%
\begin{eqnarray}
162b_{0}^{3}\Delta J_{\text{t}}^{\text{I}}
&=&-12b_{0}^{4}s_{0}-18b_{1}^{2}s_{0}+18b_{0}b_{2}s_{0}+8b_{0}^{2}s_{0}^{2}-9b_{1}s_{0}^{2}-s_{0}^{3}-18b_{0}^{3}s_{1}
\notag \\
&&+18b_{0}b_{1}s_{1}+9b_{0}s_{0}s_{1}-18b_{0}^{2}s_{2},  \notag \\
162b_{0}^{3}\Delta P_{\text{t}}^{\text{I}} &=&3(200-168\gamma _{M}+36\gamma
_{M}^{2}-3\pi ^{2})s_{0}b_{0}^{4}+36(7-3\gamma
_{M})s_{1}b_{0}^{3}+36s_{2}b_{0}^{2}  \notag \\
&&+4(41-18\gamma
_{M})s_{0}^{2}b_{0}^{2}-36b_{0}b_{2}s_{0}+36(s_{0}b_{1}-b_{0}s_{1})(b_{1}-s_{0})+8s_{0}^{3},
\label{dIt}
\end{eqnarray}%
for the tensorial sigma (\ref{sigmat}) and%
\begin{eqnarray}
{1458b_{0}^{4}}\Delta \tilde{J}_{\text{s}}^{\text{I}} &=&-54b_{0}^{5}\bar{s}%
+243b_{0}\bar{s}%
(b_{0}b_{2}-b_{1}^{2})+243b_{0}^{2}(b_{1}s_{0}-b_{0}s_{1})+81b_{1}(b_{1}-b_{0}^{2})%
\bar{s}^{2}  \notag \\
&&+18b_{0}\bar{s}^{2}(b_{1}\bar{s}-b_{0}s_{0})-162b_{0}s_{0}(b_{0}^{3}+b_{1}%
\bar{s})+27b_{0}^{2}s_{0}(3s_{0}+4b_{0}\bar{s})  \notag \\
&&+b_{0}^{2}\bar{s}^{2}(45b_{0}^{2}-12b_{0}\bar{s}+\bar{s}^{2}),  \label{dIs}
\end{eqnarray}%
for the scalar sigma (\ref{sigmaes}).

For class II we find%
\begin{equation}
162b_{1}^{3}\Delta \tilde{J}_{\text{t}}^{\text{II}%
}=9b_{1}(2b_{4}s_{0}-2b_{1}^{2}s_{1}+2b_{3}s_{1}+s_{1}s_{2})-s_{1}(9b_{3}s_{0}+b_{1}s_{0}^{2}+s_{1}^{2})+2b_{1}^{2}(5s_{0}s_{1}-9s_{3}),
\label{dIIt}
\end{equation}%
for the tensorial sigma.


\begin{thebibliography}{99}
\bibitem{englert} R. Brout, F. Englert and E. Gunzig, The creation of the
universe as a quantum

phenomenon, \href{https://doi.org/10.1016/0003-4916(78)90176-8}{Annals Phys.
115 (1978) 78}.

\bibitem{starobinsky} A.A. Starobinsky, A new type of isotropic cosmological
models without singularity, \href{https://doi.org/10.1016/0370-2693(80)90670-X/037026938090670X}%
{\href{https://doi.org/10.1016/0370-2693(80)90670-X}{Phys. Lett. B 91 (1980)
99}}.

\bibitem{kazanas} D. Kazanas, Dynamics of the universe and spontaneous
symmetry breaking, \href{https://doi.org/10.1086/183361}{Astrophys. J. 241
(1980) L59}.

\bibitem{sato} K. Sato, First-order phase transition of a vacuum and the
expansion of the universe, Monthly Notices of the Royal Astr. Soc. 195
(1981) 467.

\bibitem{guth} A.H. Guth, Inflationary universe: A possible solution to the
horizon and flatness problems, \href{https://doi.org/10.1103/PhysRevD.23.347}%
{Phys. Rev. D23 (1981) 347}.

\bibitem{linde} A.D. Linde, A new inflationary universe scenario: A possible
solution of the horizon, flatness, homogeneity, isotropy and primordial
monopole problems, \href{https://doi.org/10.1016/0370-2693(82)91219-9}{Phys.
Lett. B108 (1982) 389}.

\bibitem{steinhardt} A. Albrecht and P.J. Steinhardt, Cosmology for grand
unified theories with radiatively induced symmetry breaking, \href{https://doi.org/10.1103/PhysRevLett.48.1220}%
{Phys. Rev. Lett. 48 (1982) 1220}.

\bibitem{linde2} A.D. Linde, Chaotic inflation, \href{https://doi.org/10.1016/0370-2693(83)90837-7}%
{Phys. Lett. B129 (1983) 177}.

\bibitem{mukh} V.F. Mukhanov, G.V. Chibisov, Quantum fluctuations and a
nonsingular universe, \href{http://www.jetpletters.ac.ru/ps/1510/article_23079.shtml}%
{JETP Lett. 33 (1981) 532}, Pisma Zh. Eksp. Teor. Fiz. 33 (1981) 549.

\bibitem{mukh2} V.F. Mukhanov and G. Chibisov, The Vacuum energy and large
scale structure of the universe, \href{http://jetp.ac.ru/cgi-bin/dn/e_056_02_0258.pdf}%
{Sov. Phys. JETP 56 (1982) 258}.

\bibitem{hawk} S. Hawking, The development of irregularities in a single
bubble inflationary universe, \href{https://doi.org/10.1016/0370-2693(82)90373-2}%
{Phys. Lett. B115 (1982) 295}.

\bibitem{guth2} A.H. Guth and S. Pi, Fluctuations in the new inflationary
universe, \href{https://doi.org/10.1103/PhysRevLett.49.1110}{Phys. Rev.
Lett. 49 (1982) 1110}.

\bibitem{staro2} A.A. Starobinsky, Dynamics of phase transition in the new
inflationary universe scenario and generation of perturbations, \href{https://doi.org/10.1016/0370-2693(82)90541-X}%
{Phys. Lett. B117 (1982) 175}.

\bibitem{bardeen} J.M. Bardeen, P.J. Steinhardt and M.S. Turner, Spontaneous
creation of almost scale-free density perturbations in an inflationary
universe, \href{https://doi.org/10.1103/PhysRevD.28.679}{Phys. Rev. D28
(1983) 679}.

\bibitem{mukh3} V.F. Mukhanov, Gravitational instability of the universe
filled with a scalar

field JETP Lett. 41 (1985) 493.

\bibitem{reviews} V.F. Mukhanov, H.A. Feldman and R.H. Brandenberger, \href{https://doi.org/10.1016/0370-1573(92)90044-Z}%
{Phys. Rept. 215 (1992) 203};

S. Weinberg, \textit{Cosmology}, Oxford University Press, 2008;

A. De Felice and S. Tsujikawa, $f(R)$ theories, \href{https://doi.org/10.12942/lrr-2010-3}%
{Living Rev. Rel. 13 (2010) 3} and \href{https://arxiv.org/abs/1002.4928}{%
arXiv:1002.4928} [gr-qc].

\bibitem{encicl} J. Martin, C. Ringeval and V. Vennin, Encyclopaedia
Inflationaris, \href{https://doi.org/10.1016/j.dark.2014.01.003}{Phys. Dark
Univ. 5 (2014) 75} and \href{https://arxiv.org/abs/1303.3787}{arXiv:1303.3787%
} [astro-ph.CO];

J. Martin, C. Ringeval, R. Trotta and V. Vennin, The best inflationary
models after Planck, \href{https://doi.org/10.1088/1475-7516/2014/03/039}{%
JCAP 1403 (2014) 039} and \href{https://arxiv.org/abs/1312.3529}{%
arXiv:1312.3529} [astro-ph.CO].

\bibitem{Planck18} Planck collaboration, Planck 2018 results. X. Constraints
on inflation, \href{http://arxiv.org/abs/1807.06211}{arXiv:1807.06211}
[astro-ph.CO].

\bibitem{defelice} A. De Felice and S. Tsujikawa, $f(R)$ theories, \href{https://doi.org/10.12942/lrr-2010-3}%
{Living Rev. Rel. 13 (2010) 3} and \href{https://arxiv.org/abs/1002.4928}{%
arXiv:1002.4928} [gr-qc].

\bibitem{otherfR} T.P. Sotiriou and V. Faraoni, $f(R)$ theories of gravity, 
\href{https://doi.org/10.1103/RevModPhys.82.451}{Rev. Mod. Phys. 82 (2010)
451} \href{http://arxiv.org/abs/0805.1726}{{arXiv:0805.1726}} {[gr-qc].}

\bibitem{CMBrunning} D. Anselmi, Cosmic inflation as a renormalization-group
flow: the running of power spectra in quantum gravity, \href{https://dx.doi.org/10.1088/1475-7516/2021/01/048}%
{J. Cosmol. Astropart.} \href{https://dx.doi.org/10.1088/1475-7516/2021/01/048}%
{Phys. 01 (2021) 048}, \href{http://renormalization.com/20a4/}{20A4
Renormalization.com} and \href{https://arxiv.org/abs/2007.15023}{%
arXiv:2007.15023} [hep-th].

\bibitem{FakeOnScalar} D. Anselmi, High-order corrections to inflationary
perturbation spectra in quantum gravity, \href{https://dx.doi.org/10.1088/1475-7516/2021/02/029}%
{J. Cosmol. Astropart. Phys. 02 (2021) 029}, \href{http://renormalization.com/20a5/}%
{20A5 Renormalization.com} and \href{https://arxiv.org/abs/2010.04739}{%
arXiv:2010.04739} [hep-th].

\bibitem{double} D. Anselmi, Perturbation spectra and renormalization-group
techniques in double-field inflation and quantum gravity cosmology, \href{https://dx.doi.org/10.1088/1475-7516/2021/07/037}%
{J. Cosmol. Astropart. Phys. 07 (2021) 087}, \href{http://renormalization.com/21a2/}%
{21A2 Renormalization.com} and \href{https://arxiv.org/abs/2105.05864}{%
arXiv:2105.05864} [hep-th].

\bibitem{peters} See for example, P. Peter and J.P. Uzan, \textit{Primordial
cosmology}, Oxford University Press, 2009.

\bibitem{stelle} K.S. Stelle, Renormalization of higher derivative quantum
gravity, \href{https://doi.org/10.1103/PhysRevD.16.953}{Phys. Rev. D 16
(1977) 953}.

\bibitem{LWgrav} D. Anselmi, On the quantum field theory of the
gravitational interactions, \href{http://dx.doi.org/doi:10.1007/JHEP06(2017)086}%
{J. High Energy Phys. 06 (2017) 086}, \href{http://renormalization.com/17a3/}%
{17A3 Renormalization.com} and \href{http://arxiv.org/abs/1704.07728}{arXiv:}
\href{http://arxiv.org/abs/1704.07728}{1704.07728} [hep-th].

\bibitem{UVQG} D. Anselmi and M. Piva, The ultraviolet behavior of quantum
gravity, \href{http://dx.doi.org/10.1007/JHEP05(2018)027}{J. High Energ.
Phys. 05 (2018) 27}, \href{http://renormalization.com/18a2/}{18A2
Renormalization.com} and \href{http://arxiv.org/abs/1803.07777}{%
arXiv:1803.07777} [hep-th].

\bibitem{Absograv} D. Anselmi and M. Piva, Quantum gravity, fakeons and
microcausality, J. High Energy Phys. 11 (2018) 21, \href{http://renormalization.com/18a3/}%
{18A3 Renormalization.com} and \href{http://arxiv.org/abs/1806.03605}{%
arXiv:1806.03605} [hep-th].

\bibitem{fakeons} D. Anselmi, Fakeons and Lee-Wick models, \href{http://dx.doi.org/10.1007/JHEP02(2018)141}%
{J. High Energy Phys. 02 (2018) 141}, \href{http://renormalization.com/18a1/}%
{18A1 Renormalization.com} and \href{http://arxiv.org/abs/1801.00915}{%
arXiv:1801.00915} [hep-th].

\bibitem{ABP} D. Anselmi, E. Bianchi and M. Piva, Predictions of quantum
gravity in inflationary cosmology: effects of the Weyl-squared term, \href{https://doi.org/10.1007/JHEP07(2020)211}%
{J. High Energy Phys. 07 (2020) 211}, \href{http://renormalization.com/20a2/}%
{20A2 Renormalization.com} and \href{http://arxiv.org/abs/2005.10293}{%
arXiv:2005.10293} [hep-th].

\bibitem{pieroni} P. Bin\'{e}truy, E. Kiritsis, J. Mabillard, M. Pieroni and
C. Rosset, Universality classes for models of inflation, \href{https://doi.org/10.1088/1475-7516/2015/04/033}%
{J. Cosmol. Astropart. Phys. 04 (2015) 033}\ and \href{https://arxiv.org/abs/1407.0820}%
{arXiv:1407.0820} [astro-ph.CO];

see also P. Bin\'{e}truy, J. Mabillard and M. Pieroni, Universality in
generalized models of inflation, \href{https://doi.org/10.1088/1475-7516/2017/03/060}%
{J. Cosmol. Astropart. Phys. 03 (2017) 060} and \href{https://arxiv.org/abs/1611.07019v2}%
{arXiv:1611.07019} [gr-qc].

\bibitem{kiritsis} E. Kiritsis, Asymptotic freedom, asymptotic flatness and
cosmology, \href{https://doi.org/10.1088/1475-7516/2013/11/011}{J. Cosmol.
Astropart. Phys.} \href{https://doi.org/10.1088/1475-7516/2013/11/011}{11
(2013) 011} and \href{https://arxiv.org/abs/1307.5873}{arXiv:1307.5873}
[hep-th].

\bibitem{run1} A. Kosowsky and M.S. Turner, CBR anisotropy and the running
of the scalar spectral index, \href{https://doi.org/10.1103/PhysRevD.52.R1739}%
{Phys. Rev. D52 (1995) 1739} and \href{https://arxiv.org/abs/astro-ph/9504071}%
{arXiv:astro-ph/9504071};

D.J.H. Chung, G. Shiu and M. Trodden, Running of the scalar spectral index
from inflationary models, \href{https://doi.org/10.1103/PhysRevD.68.063501}{%
Phys. Rev. D 68 (2003) 063501} and \href{https://arxiv.org/abs/astro-ph/0305193}%
{arXiv:astro-ph/0305193};

J.E. Lidsey and R. Tavakol, Running of the scalar spectral index and
observational signatures of inflation, \href{https://doi.org/10.1016/j.physletb.2003.07.091}%
{Phys. Lett. B575 (2003) 157} and \href{https://arxiv.org/abs/astro-ph/0304113}%
{arXiv:astro-ph/0304113}.

\bibitem{run2} T.T. Nakamura and E.D. Stewart, The spectrum of cosmological
perturbations produced by a multi-component inflaton to second order in the
slow-roll approximation, \href{https://doi.org/10.1016/0370-2693(96)00594-1}{%
Phys. Lett. B381 (1996) 413} and \href{https://arxiv.org/abs/astro-ph/9604103}%
{arXiv:astro-ph/9604103};

E.D. Stewart, and J.-O. Gong, The density perturbation power spectrum to
second-order corrections in the slow-roll expansion, \href{https://doi.org/10.1016/S0370-2693(01)00616-5}%
{Phys. Lett. B 510 (2001) 1} and \href{https://arxiv.org/abs/astro-ph/0101225}%
{arXiv:astro-ph/0101225};

W.H. Kinney, Inflation: flow, fixed points and observables to arbitrary
order in slow roll, \href{https://10.1103/PhysRevD.66.083508}{Phys. Rev. D66
(2002) 083508} and \href{https://arxiv.org/abs/astro-ph/0206032}{%
arXiv:astro-ph/0206032};

T. Zhu, A. Wang, G. Cleaver, K. Kirsten and Q. Sheng, Gravitational quantum
effects on power spectra and spectral indices with higher-order corrections, 
\href{https://10.1103/PhysRevD.90.063503}{Phys. Rev. D90 (2014) 063503} and 
\href{https://arxiv.org/abs/1405.5301}{arXiv:1405.530}1 [astro-ph.CO];

M. Zarei, On the running of the spectral index to all orders: a new model
dependent approach to constrain inflationary models, \href{https://10.1088/0264-9381/33/11/115008}%
{Class. Quantum Grav. 33 (2016) 115008} and \href{https://arxiv.org/abs/1408.6467}%
{arXiv:1408.6467} [astro-ph.CO];

H. Motohashi and W. Hu, Generalized slow roll in the unified effective field
theory of inflation, \href{https://10.1103/PhysRevD.96.023502}{Phys. Rev. D
96 (2017) 023502} and \href{https://arxiv.org/abs/1704.01128}{%
arXiv:1704.01128} [hep-th];

A. Karam, T. Pappas and K. Tamvakis, Frame-dependence of higher-order
inflationary observables in scalar-tensor theories, \href{https://doi.org/10.1103/PhysRevD.96.064036}%
{Phys. Rev. D 96 (2017) 064036} and \href{https://arxiv.org/abs/1707.00984}{%
arXiv:1707.00984} [gr-qc];

G.-H. Ding, J. Qiao, Q. Wu, T. Zhu and A. Wang, Inflationary perturbation
spectra at next-to-leading slow-roll order in effective field theory of
inflation, \href{https://10.1140/epjc/s10052-019-7496-7}{Eur. Phys. J. C 79
(2019) 976} and \href{https://arxiv.org/abs/1907.13108}{arXiv:1907.13108}
[gr-qc].

\bibitem{alfat} R. Kallosh and A. Linde, Universality class in conformal
inflation, \href{https://doi.org/10.1088/1475-7516/2013/07/002}{\href{https://dx.doi.org/10.1088/1475-7516/2021/01/048}%
{J. Cosmol. Astropart.} \href{https://dx.doi.org/10.1088/1475-7516/2021/01/048}%
{Phys.}} \href{https://doi.org/10.1088/1475-7516/2013/07/002}{07 (2013) 002}%
, \href{https://arxiv.org/abs/1306.5220}{arXiv:1306.5220} [hep-th];

R. Kallosh and A. Linde, Planck, LHC, and $\alpha $-attractors, \href{https://doi.org/10.1103/PhysRevD.91.083528}%
{Phys. Rev. D 91 (2015) 083528} and \href{https://arxiv.org/abs/1502.07733}{%
arXiv:1502.07733} [astro-ph.CO].

\bibitem{Palatini} F. Bauer and D. A. Demir, Inflation with non-minimal
coupling: metric vs. Palatini formulations, \href{https://doi.org/10.1016/j.physletb.2008.06.014}%
{Phys. Lett. B 665 (2008) 222} and \href{https://arxiv.org/abs/0803.2664}{%
arXiv:0803.2664} [hep-ph].

\bibitem{expinfl} Y. N. Obukov, Spin-driven inflation, \href{https://doi.org/10.1016/0375-9601(93)91059-E}%
{Phys. Lett. A 182 (1993) 214} and \href{https://arxiv.org/abs/gr-qc/0008015}%
{arXiv:0008015} [gr-qc];

E. D. Stewart, Inflation, Supergravity and Superstrings, \href{https://doi.org/10.1103/PhysRevD.51.6847}%
{Phys. Rev. D 51 (1995) 6847} and \href{https://arxiv.org/abs/hep-ph/9405389}%
{arXiv:9405389} [hep-ph];

G. Dvali and S.-H. Henry Tye, Brane Inflation, \href{https://doi.org/10.1016/S0370-2693(99)00132-X}%
{Phys. Lett. B 450 (1999) 72} and \href{https://arxiv.org/abs/hep-ph/9812483}%
{arXiv:9812483} [hep-ph];

M. Cicoli, C. Burgess, and F. Quevedo, Fibre inflation: observable gravity
waves from IIB string compactifications, \href{https://doi.org/10.1088/1475-7516/2009/03/013}%
{J. Cosmol. Astropart. Phys. 0903 (2009) 013} and \href{https://arxiv.org/abs/0808.0691}%
{arXiv:0808.0691};

G.F. Giudice and H.M. Lee, Unitarizing Higgs inflation, \href{https://doi.org/10.1016/j.physletb.2010.10.035}%
{Phys. Lett. B 694 (2011) 294} and \href{https://arxiv.org/abs/1010.1417}{%
arXiv:1010.1417}.

\bibitem{mutatedhilltop} B. Kumar Pal, Supratik Pal, B. Basu, Mutated
Hilltop inflation: a natural choice for early universe, \href{https://doi.org/10.1088/1475-7516/2010/01/029}%
{J. Cosmol. Astropart. Phys. 1001 (2010) 029} and \href{https://arxiv.org/abs/0908.2302}%
{arXiv:0908.2302} [hep-th].

\bibitem{LFI} A. D. Linde, Chaotic inflating universe, JETP Lett. 38 (1983)
176;

M. Madsen and P. Coles, Chaotic inflation, \href{https://doi.org/10.1016/0550-3213(88)90004-1}%
{Nucl. Phys. B 298 701 (1988) 701};

G. Lazarides and Q. Shafi, A predictive inflationary scenario without the
gauge singlet, \href{https://doi.org/10.1016/0370-2693(93)90595-9}{Phys.
Lett. B 308 (1993) 17} and \href{https://arxiv.org/abs/hep-ph/9304247}{%
arXiv:hep-ph/9304247};

L. Kofman, A.D. Linde and A.A. Starobinsky, Reheating after inflation, \href{https://doi.org/10.1103/PhysRevLett.73.3195}%
{Phys. Rev. Lett. 73 (1994) 3195} and \href{https://arxiv.org/abs/hep-th/9405187}%
{arXiv:hep-th/9405187};

G. Lazarides and Q. Shafi, Topological defects and inflation,\href{https://doi.org/10.1016/0370-2693(96)00136-0}%
{ Phys. Lett. B 372 (1996) 20} and \href{https://arxiv.org/abs/hep-ph/9510275}%
{arXiv:hep-ph/9510275}.

\bibitem{DWI} P. Peter, Spontaneous current generation in cosmic strings, 
\href{https://doi.org/10.1103/PhysRevD.49.5052}{Phys. Rev. D 49 (1994) 5052}
and \href{https://arxiv.org/abs/hep-ph/9312280}{arXiv:hep-ph/9312280}.

\bibitem{MSSMI} ~R. Allahverdi, K. Enqvist, J. Garcia-Bellido and A.
Mazumdar, Gauge invariant MSSM inflaton, \href{https://doi.org/10.1103/PhysRevLett.97.191304}%
{Phys. Rev. Lett. 97 (2006) 191304} and \href{https://arxiv.org/abs/hep-ph/0605035}%
{arXiv:hep-ph/0605035}.

\bibitem{baumann} D. Baumann, TASI lectures on inflation, \href{https://arxiv.org/abs/0907.5424}%
{arXiv:0907.5424} [hep-th].

\bibitem{radi} V. N. Senoguz and Q. Shafi, Chaotic inflation, radiative
corrections and precision cosmology, \href{https://doi.org/10.1016/j.physletb.2008.08.017}%
{Phys. Lett. B668 (2008) 6} and \href{https://arxiv.org/abs/0806.2798}{%
arXiv:0806.2798};

A. Barvinsky, A.Y. Kamenshchik and A. Starobinsky, Inflation scenario via
the Standard Model Higgs boson and LHC, \href{https://doi.org/10.1088/1475-7516/2008/11/021}%
{J. Cosmol. Astropart. Phys. 0811 (2008) 021} and \href{https://arxiv.org/abs/0809.2104}%
{arXiv:0809.2104}.

\bibitem{loop} P. Binetruy and G. Dvali, D term inflation, \href{https://doi.org/10.1016/S0370-2693(96)01083-0}%
{Phys. Lett. B 388 (1996) 241} and \href{https://arxiv.org/abs/hep-ph/9606342}%
{arXiv:hep-ph/9606342}.

\bibitem{nat} K. Freese, J.A. Frieman and A. V. Olinto, Natural inflation
with pseudo - Nambu-Goldstone bosons, \href{https://doi.org/10.1103/PhysRevLett.65.3233}%
{Phys. Rev. Lett. 65 (1990) 3233};

F.C. Adams, J.R. Bond, K. Freese, J.A. Frieman and A.V. Olinto, Natural
inflation: Particle physics models, power law spectra for large scale
structure, and constraints from COBE, \href{https://doi.org/10.1103/PhysRevD.47.426}%
{Phys. Rev. D47 (1993) 426} and \href{https://arxiv.org/abs/hep-ph/9207245}{%
arXiv:hep-ph/9207245}.

\bibitem{PLI} L. Abbott and M.B. Wise, Constraints on generalized
inflationary cosmologies, \href{https://doi.org/10.1016/0550-3213(84)90329-8}%
{Nucl. Phys. B 244 (1984) 541}.

\bibitem{RGI} M. Fairbairn, L. Lopez Honorez and M. Tytgat, Radion assisted
gauge inflation, \href{https://doi.org/10.1103/PhysRevD.67.101302}{Phys.
Rev. D 67 (2003) 101302} and \href{https://arxiv.org/abs/hep-ph/0302160}{%
arXiv:hep-ph/0302160}.

\bibitem{Litebird} M. Hazumi, at al., LiteBIRD satellite: JAXA's new
strategic L-class mission for all-sky surveys of cosmic microwave background
polarization, Proceedings Volume 11443, Space Telescopes and Instrumentation
2020: Optical, Infrared, and Millimeter Wave, 114432F (2020), \href{https://arxiv.org/abs/2101.12449}%
{arXiv:2101.12449} [astro-ph.IM].

\bibitem{salvio} T. Clunan and M. Sasaki, Tensor ghosts in the inflationary
cosmology, \href{https://doi.org/10.1088/0264-9381/27/16/165014}{Class.
Quant. Grav. 27 (2010) 165014} and \href{http://arxiv.org/abs/0907.3868}{%
arXiv:0907.3868} [hep-th];

N. Deruelle, M. Sasaki, Y. Sendouda and A. Youssef, Inflation with a Weyl
term, or ghosts at work, \href{https://doi.org/10.1088/1475-7516/2011/03/040}%
{JCAP 1103 (2011) 040} and \href{https://arxiv.org/abs/1012.5202v1}{%
arXiv:1012.5202} [gr-qc];

N. Deruelle, M. Sasaki, Y. Sendouda and A. Youssef, Lorentz-violating vs
ghost gravitons: the example of Weyl gravity, \href{https://doi.org/10.1007/JHEP09(2012)009}%
{J. High Energ. Phys. 2012 (2012) 9} and \href{http://arxiv.org/abs/1202.3131}%
{arXiv:1202.3131} [gr-qc];

C. Fang and QG. Huang, The trouble with asymptotically safe inflation, \href{http://dx.doi.org/10.1140/epjc/s10052-013-2401-2}%
{Eur. Phys. J. C 73, 2401 (2013)} and \href{https://arxiv.org/abs/1210.7596}{%
arXiv:1210.7596} [hep-th];

Y.S. Myung and T. Moon, Primordial massive gravitational waves from
Einstein-Chern-Simons-Weyl gravity, \href{https://doi.org/10.1088/1475-7516/2014/08/061}%
{JCAP 08 (2014) 061} and \href{http://arxiv.org/abs/1406.4367}{%
arXiv:1406.4367} [gr-qc];

K. Kannike, G. H\"{u}tsi, L. Pizza, A. Racioppi, M. Raidal, A. Salvio and A.
Strumia, Dynamically induced Planck scale and inflation, \href{https://doi.org/10.1007/JHEP05(2015)065}%
{J. High Energy Phys. 05 (2015) 065} and \href{https://arxiv.org/abs/1502.01334}%
{arXiv:1502.01334} [astro-ph.CO];

M.M. Ivanov and A.A. Tokareva, Cosmology with a light ghost, \href{https://doi.org/10.1088/1475-7516/2016/12/018}%
{JCAP 12 (2016) 018} and \href{https://arxiv.org/abs/1610.05330}{%
arXiv:1610.05330} [hep-th];

A. Salvio, Inflationary perturbations in no-scale theories, \href{https://doi.org/10.1140/epjc/s10052-017-4825-6}%
{Eur. Phys. J. C77 (2017)} \href{https://doi.org/10.1140/epjc/s10052-017-4825-6}%
{267} and \href{https://arxiv.org/abs/1703.08012}{arXiv:1703.08012}
[astro-ph.CO].
\end{thebibliography}
\end{document}